\documentclass[fleqn,usenatbib]{mnras}

\usepackage{newtxtext,newtxmath}
\usepackage{chemformula}
\usepackage[T1]{fontenc}

\DeclareRobustCommand{\VAN}[3]{#2}
\let\VANthebibliography\thebibliography
\def\thebibliography{\DeclareRobustCommand{\VAN}[3]{##3}\VANthebibliography}

\usepackage{graphicx}	% Including figure files
\usepackage{amsmath}	% Advanced maths commands

\usepackage{comment}
\usepackage{multirow}
\usepackage{hyperref}

\newcommand{\orcid}[1]{\href{https://orcid.org/#1}{\hskip2pt\includegraphics[width=9pt]{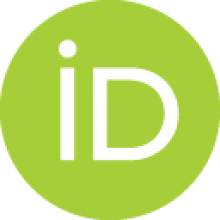}}}
\title[Acetaldehyde formation]{Quantum mechanical modeling of the grain-surface formation of acetaldehyde on H$_2$O:CO dirty ice surfaces}

\author[J. Perrero et al.]{
Jessica Perrero\orcid{0000-0003-2161-9120},$^{1,2}$
Piero Ugliengo\orcid{0000-0001-8886-9832},$^{2}$ \thanks{E-mail: piero.ugliengo@unito.it}
Cecilia Ceccarelli\orcid{0000-0001-9664-6292},$^{3}$
and Albert Rimola\orcid{0000-0002-9637-4554}$^{1}$
\thanks{E-mail: albert.rimola@uab.cat}
\\
$^{1}$Departament de Qu\'{i}mica, Universitat Aut\`{o}noma de Barcelona, Bellaterra, 08193, Catalonia, Spain\\
$^{2}$Dipartimento di Chimica and Nanostructured Interfaces and Surfaces (NIS) Centre, Universit\`{a} degli Studi di Torino, via P. Giuria 7, 10125, Torino, Italy\\
$^{3}$Univ. Grenoble Alpes, CNRS, Institut de Plan\'{e}tologie et d'Astrophysique de Grenoble (IPAG), 38000 Grenoble, France
}

\date{Accepted XXX. Received YYY; in original form ZZZ}

\pubyear{2023}

\begin{document}
\label{firstpage}
\pagerange{\pageref{firstpage}--\pageref{lastpage}}
\maketitle

% Abstract of the paper
\begin{abstract}
Acetaldehyde (CH$_3$CHO) is one of the most detected interstellar Complex Organic Molecule (iCOM) in the interstellar medium (ISM).  
These species  have a potential biological relevance, as they can be precursors of more complex species from which life could have emerged.
The formation of iCOMs in the ISM is a challenge  and a matter of debate, whether gas-phase, grain-surface chemistry or both are needed for their synthesis. 
In the gas-phase, CH$_3$CHO can be efficiently synthesized from ethanol and/or ethyl radical. 
On the grain-surfaces, radical-radical recombinations were traditionally invoked.
However, several pitfalls have been recently identified, such as the presence of energy barriers and competitive side reactions (i.e., H abstractions). 
Here we investigate a new grain-surface reaction pathway for the formation of acetaldehyde, namely the reaction between CH$_3$ and a CO molecule of a dirty water/CO ice followed by hydrogenation of its product, CH$_3$CO.
To this end, we carried out \textit{ab initio} computations of the reaction occurring on an ice composed by 75\% water and 25\% CO molecules. 
We found that the CH$_3$ + CO$_{(ice)}$ reaction exhibits barriers difficult to overcome in the ISM, either adopting a Langmuir-Hinshelwood or an Eley-Rideal mechanism. 
The subsequent hydrogenation step is found to be barrierless, provided that the two reacting species have the correct orientation. 
Therefore, this pathway seems unlikely to occur in the ISM.

\end{abstract}

% Select between one and six entries from the list of approved keywords.
% Don't make up new ones.

%%%%%%%%%%%%%%%%%%%%%%%%%%%%%%%%%%%%%%%%%%%%%%%%%%

%%%%%%%%%%%%%%%%% BODY OF PAPER %%%%%%%%%%%%%%%%%%

\section{Introduction}

%CH3CHO and iCOMs
Acetaldehyde is one of the most common detected interstellar Complex Organic Molecules (iCOMs), which are chemical compounds defined as molecules with six or more atoms that contain at least one carbon atom \citep{Herbst2009,Herbst2017,Ceccarelli2017}. 
Since their detection in star-forming regions, iCOMs rose the interest of scientists. 
There is evidence that some iCOMs formed in the Interstellar Medium (ISM) were inherited by the small objects of the Solar System \citep{Astrochem-Heritage_2012,Ceccarelli_2014,cazaux2003,alma-pils2018,Bianchi2019,Drozdovskaya2019}.
These species, after thermal and hydrothermal alterations, can be converted into more complex organic species \citep{Alexander2014,Rotelli2016,Yabuta2007,Callahan2011}, therefore potentially paving the way to the emergence of life on Earth.

%detection
Acetaldehyde was first detected in Sagittarius B2 by  \cite{gottlieb1973}, \cite{foukiris1974} and \cite{gilmore1976}. Some years later, it was also detected in cold clouds, TMC-1 and L134N, by \cite{matthews1985}. 
This molecule has been found in a large number of environments: 
cold prestellar cores \citep{bacmann2012,Scibelli2020},
hot cores \citep{blake1987,law2021}, 
hot corinos \citep{cazaux2003,chahine2022}, 
protostellar molecular shocks \citep{lefloch2017, desimone2020}, 
young discs \citep{codella2018,lee2019} 
and also in comets \citep{crovisier2004,biver2021}.

%formation
The presence of acetaldehyde (CH$_3$CHO) and other iCOMs in cold prestellar cores demonstrates that their synthesis cannot be the result of grain-surface reactions involving migration of species (i.e. radicals) other than H and O \citep[e.g.][]{bacmann2012, Ceccarelli2022-PP7}.

%paradigms
The reaction pathways that lead to iCOMs are still matter of debate, as both gas-phase and grain-surface chemistry are invoked to play a crucial role in their synthesis \citep[e.g.][]{garrod2008,Balucani2015,Ceccarelli2022-PP7}.
Several paradigms for the formation of iCOMs were proposed in the literature, the most popular being schemes based on gas-phase reactions \citep[e.g.][]{charnley1992,charnley1997,Taquet2016,Balucani2015,skouteris2018,vazart2020} and a network of radical-radical couplings occurring on the surface of grains \citep[e.g.][]{garrod2006, garrod2008, jin_garrod_2020}. 
Other paradigms include a mechanism based on the condensation of atomic C \citep{Ruaud2015, Krasnokutski2020}, on the excited O-atom insertion \citep{Bergner2017_Oins, Bergner2019}, or on the formation of HCO radical on ice surfaces as a parent precursor of other iCOMs \citep{Fedoseev2015, Simons2020}. 

%previous studies: gas, theoretical
For what concerns acetaldehyde, several works investigated its formation routes from an experimental and theoretical point of view.
In the gas-phase, the reaction between ethyl radical and atomic oxygen was proposed to yield CH$_3$CHO by \cite{charnley2004}, as well as the insertion of CH into methanol \citep{vasyunin2017}. 
In KIDA \citep{wakelam2012} and UMIST \citep{mcelroy2013} databases, an ionic route involving protonated acetaldehyde (obtained from dimethyl ether) is present. 
Finally, in 2018, the idea that acetaldehyde (and other iCOMs) can arise from the chemical transformation of ethanol in the gas-phase was proposed \citep{skouteris2018}. 
In this chemical network, called the genealogical tree of ethanol, the latter is the precursor (the parent molecule) that give rise to different iCOMs such as glycolaldehyde, acetic acid, formic acid and acetaldehyde, among others. 
More recently, \citet{vazart2020} carried out a systematic study of all gas-phase reactions present in the literature, that lead to the formation of acetaldehyde, and performed new \textit{ab initio} computations for reactions having only guessed product and rate constants.
These authors confirmed the pathways starting from ethyl radical \citep{charnley2004} and ethanol \citep{skouteris2018} and discarded the others (as they used incorrect product or rate constants).
In their study, \citet{vazart2020} also showed that the ethanol genealogical tree is currently the most promising explanation for the synthesis of acetaldehyde in warm objects.
\citet{perrero2022ethanol} proposed that ethanol would be formed on the grain-surface by the reaction of CCH with a water molecule of the ice, followed by hydrogenation of the produced vinyl alcohol.
Remarkably, the presence of frozen ethanol has been tentatively detected by James Web Space Telescope (JWST) observations (although needs confirmation) \citep{McClure2023-JWSTices, Yang2022-JWST}, supporting the hypothesis of the ethanol being the mother of acetaldehyde.

%previous studies: mantle, experimental
On the grain surfaces, experimental results are in some cases contradictory. 
Acetaldehyde was formed in ices containing H$_2$O, CO, CH$_4$ and CH$_3$OH processed by UV-irradiation \citep{Moore_Hudson_1998,Bennett2005,oberg2009,oberg2010,paardekooper2016,martindom2020}. 
In the experiments by \cite{chuang2021}, a number of iCOMs  were obtained by irradiating with 200 keV H$^+$ ions C$_2$H$_2$:H$_2$O ices, including acetaldehyde.
However, in experiments using a matrix isolation technique of UV-illuminated methanol ices, where the presence of radicals could be monitored, \cite{gutierrez2021} found that several iCOMs were formed, except acetaldehyde. 
Finally, the experiments by \cite{Fedoseev:2022} produced acetaldehyde and its precursor, ketene, via a non-energetic pathway, where CO is co-deposited with C, H and H$_2$O at 10 K. 

%previous studies: mantle, theoretical
From a theoretical point of view, the radical-radical coupling of HCO and CH$_3$ was proposed by \cite{garrod2006} to yield CH$_3$CHO on icy grain surfaces, as it was supposed that the reaction is barrierless due to taking place between two radical species. 
However, successive studies showed that not only this reaction has an energy barrier (because the radicals are physisorbed on the ice surface and have to break the intermolecular forces with the surface to react), but it also presents a competitive channel leading to the formation of CH$_4$ + CO \citep{enrique-romero2019,enrique-romero2020}. 
The same reaction simulated on a model of CO ice gave similar results, culminating in one of these three outcomes: formation of CH$_3$CHO, H-abstraction yielding CH$_4$ + CO or no reaction taking place \citep{Lamberts2019}.
A recent study by \cite{enrique-romero2021} concluded that the efficiency of acetaldehyde formation via radical coupling is a strong function of the mobility of the radicals on the grain surface (and, consequently, of the grain temperature), in which the easier the diffusion of HCO and CH$_3$, the less acetaldehyde formation efficiency.
Finally, recent kinetic calculations of \cite{chouika2022} on the reaction between atomic carbon and methanol proposed by \cite{singh2019} shows the presence of a slow reaction step preceding the barrierless formation of acetaldehyde from the radicals HCO and CH$_3$. 
However, at low temperature, the product can be formed thanks to tunnelling effects.
We notice, however, that the simultaneous presence of methanol and atomic carbon on the grain-surfaces is unlike, as methanol is mostly formed by the hydrogenation of frozen CO, i.e. when carbon is prevalently locked into CO.

%our work
In summary, there is evidence that, in hot cores/corinos, acetaldehyde is formed in the gas-phase by reactions occurring in the warm gas, but it is still unclear whether this is the only mechanism at work, especially in cold environments.
In light of this situation, in this work, we decided to delve into the non-energetic formation of acetaldehyde on the surface of dust grains, when radicals cannot diffuse.

When dealing with surface reactions, several mechanisms can operate: 
i) Langmuir–Hinshelwood \citep[LH:][]{langmuir1922,hinshelwood1930} reactions, which are efficient in the case that one of the reactants can easily diffuse on the surface of the grains \citep{1993HasegawaHerbst}; 
ii) Eley–Rideal \citep[ER: ][]{eley1940} reactions, in which species from the gas-phase directly react with surface molecules, avoiding diffusion, but are efficient only if there are sufficient reactive species on the grain surface and if the reactions do not possess an activation energy \citep{Ruaud2015}; 
iii) Harris-Kasemo hot atom reactions, in which a high energy species has enough energy to overcome the diffusion barriers and travels on the surface until all excess energy is lost or it reacts \citep{harris1981}; 
iv) reactions of suprathermal species generated by the excitation and/or ionization caused by cosmic ray bombardment \citep{shingledecker2018,shingledecker_2018apj,paulive:2021}.

In the present work, we investigated the acetaldehyde formation through a two-steps mechanism based on a "radical + ice" reaction, as done for the synthesis of formamide \cite[CN + H$_2$O$_{(ice)}$:][]{Rimola2018}) and of vinyl alcohol/ethanol \cite[CCH + H$_2$O$_{(ice)}$:][]{perrero2022ethanol}). 
Here, we propose the reaction of a methyl radical, CH$_3$, with a CO molecule belonging to the ice mantle of the grain, thus avoiding competitive reactions as H-abstractions, followed by a hydrogenation step:

\begin{equation}
\text{CH$_3$ + CO $\to$ CH$_3$CO}
\label{chem_eqn:acetyl}    
\end{equation}
\begin{equation}
\text{CH$_3$CO + H $\to$ CH$_3$CHO}
\label{chem_eqn:acetaldehyde}
\end{equation}

In the first step, which involves the coupling of the methyl radical CH$_3$ with CO$_{(ice)}$, we assume that either CH$_3$ is adsorbed close to the CO and diffuses to react with it (LH mechanism) or that it lands directly on the CO, reacting immediately with it (ER mechanism). 
The second step consists in the hydrogenation of the acetyl (CH$_3$CO) radical so formed, which is expected to be almost barrierless as it involves a H atom reacting with a radical, whether it diffuses on the surface (LH) or it comes from the gas-phase (ER).

This work is organized as follows: in Section \ref{sec:methods} we report the methodology (including benchmark studies), in Section \ref{sec:results} we present the results and Section \ref{sec:discuss} is dedicated to their discussion and comparison with other studies.
Section \ref{sec:conclusions} concludes the article.

\section{Methodology}\label{sec:methods}

\subsection{Computational details}

We employed CRYSTAL17 \citep{crystal} and Gaussian16 \citep{gaussian} software packages, to execute periodic and molecular calculations, respectively. CRYSTAL17, at variance with other periodic codes, adopts localized Gaussian functions as basis sets (in a similar approach to that for Gaussian16), and it works with systems that range from zero to three periodic dimensions, avoiding the fake 3D replica of surface models that would arise when working with plane waves basis sets.

To reduce the computational cost of the periodic simulations, all the structures were optimized using the approximated HF-3c method followed by a single point energy calculation at the Density Functional Theory (DFT) level, hereafter referred to as DFT//HF-3c. To check the accuracy of this scheme, we performed a benchmark study, which splits in two phases: i) we compared the performance of DFT//HF-3c (CRYSTAL17) vs CCSD(T)//DFT in the gas-phase (Gaussian16) to select the best performing DFT functional, and ii) we applied the ONIOM2 correction \citep{dapprich:1999} with the extrapolation scheme of \citet{okoshi2015} in order to compare DFT//HF-3c against ONIOM2//DFT on the surface. 

The HF-3c method was used in the most time-consuming periodic calculations, namely, geometry optimizations and frequency calculations \citep{sure:2013}. HF-3c is based on the Hartree-Fock (HF) method computed with a minimal basis set (MINIX, \citep{tatewaki:1980}, in which three empirical corrections (3c) were added: the dispersion energy D3(BJ) \citep{grimme:2010}, the basis set superposition error (BSSE) correction with the geometrical counterpoise (gCP) \citep{kruse:2012}, and the short range bond (SRB) correction to fix overestimated covalent bond lengths for electronegative elements \citep{grimme:2011,brandenburg:2013}. Subsequently, the energies were refined with single point calculations onto the HF-3c optimized structures at the desired DFT level of theory, following an approach that provided accurate results in several cases, from molecular crystals \citep{cutini:2016}, polypeptides \citep{cutini:2017}, pure-silica zeolites \citep{cutini:2019} to the computation of binding energies on crystalline and amorphous pure water ice \citep{ferrero:2020,perrero2022be}.

In the periodic calculations, all the stationary points of a potential energy surfaces (PES) were characterized by the calculation of the harmonic frequencies at $\Gamma$ point as minima (reactants, products) or first order saddle points (transition states). Each Hessian matrix element was computed numerically by means of a 3-points formula based on two displacements of 0.003 \AA~from the minimum along each Cartesian coordinate. The zero-point energy correction was computed with the standard rigid rotor/harmonic oscillator formalism \citep{mcquarrie}. Since the systems are open-shell in nature, calculations were performed within the unrestricted formalism.
The threshold parameters for the evaluation of the Coulomb and exchange bi-electronic integrals were set equal to 10$^{-7}$, 10$^{-7}$, 10$^{-7}$, 10$^{-7}$, and 10$^{-25}$. The sampling of the reciprocal space was conducted with a Pack–Monkhorst mesh \citep{pack:1977}, with a shrinking factor of 2, which generates 4k points in the first Brillouin zone.

\subsection{\ch{H2O}:CO ice surface models}

\begin{figure}
    \centering
    \includegraphics[width=0.8\columnwidth]{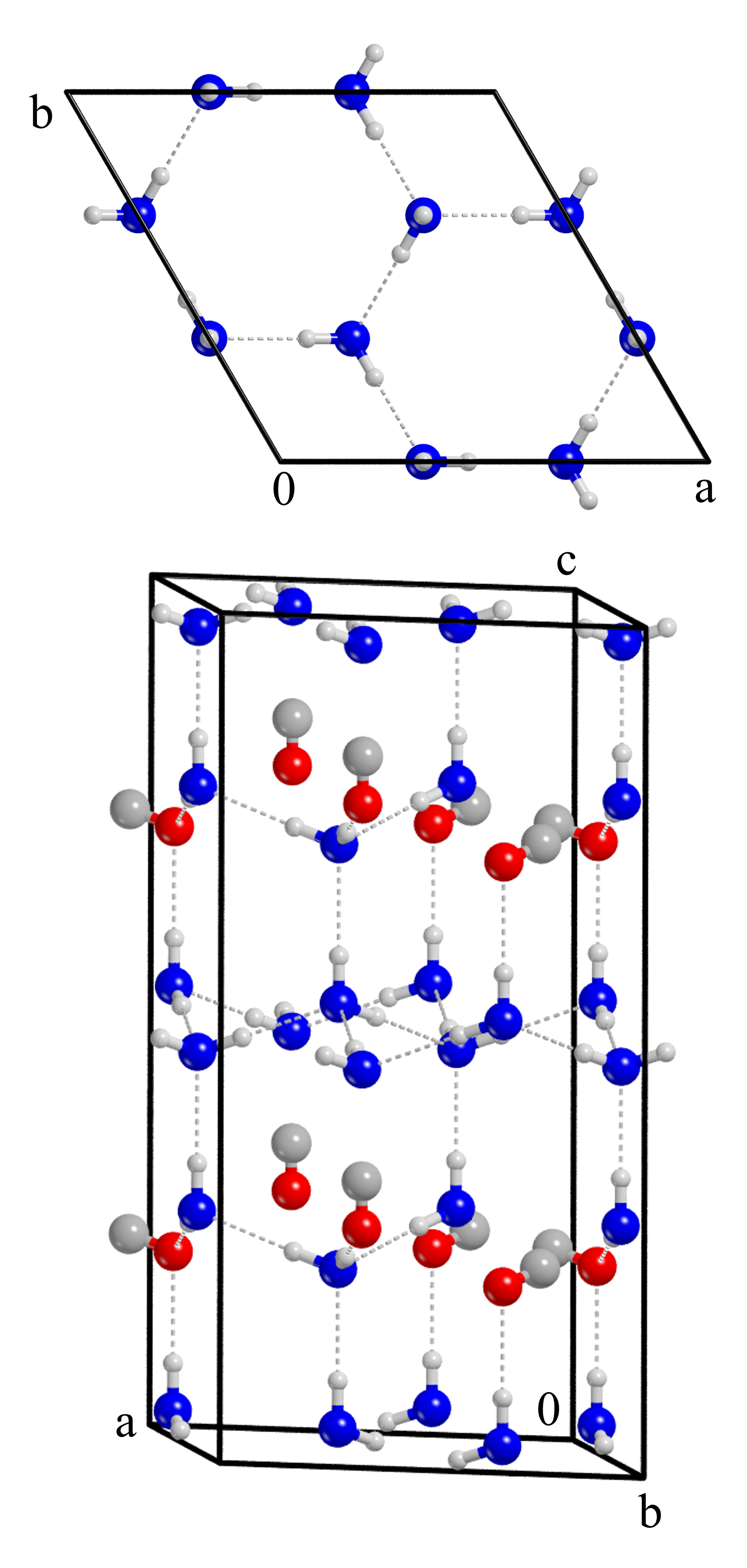}
    \caption{Top and lateral view of the dirty ice bulk model where one every four water molecules was replaced by CO. Color code: H, white; C, grey; O atoms belonging to water, blue; O atoms belonging to the CO, red.}
    \label{fig:bulk}
\end{figure}

\begin{figure}
    \centering
    \includegraphics[width=0.85\columnwidth]{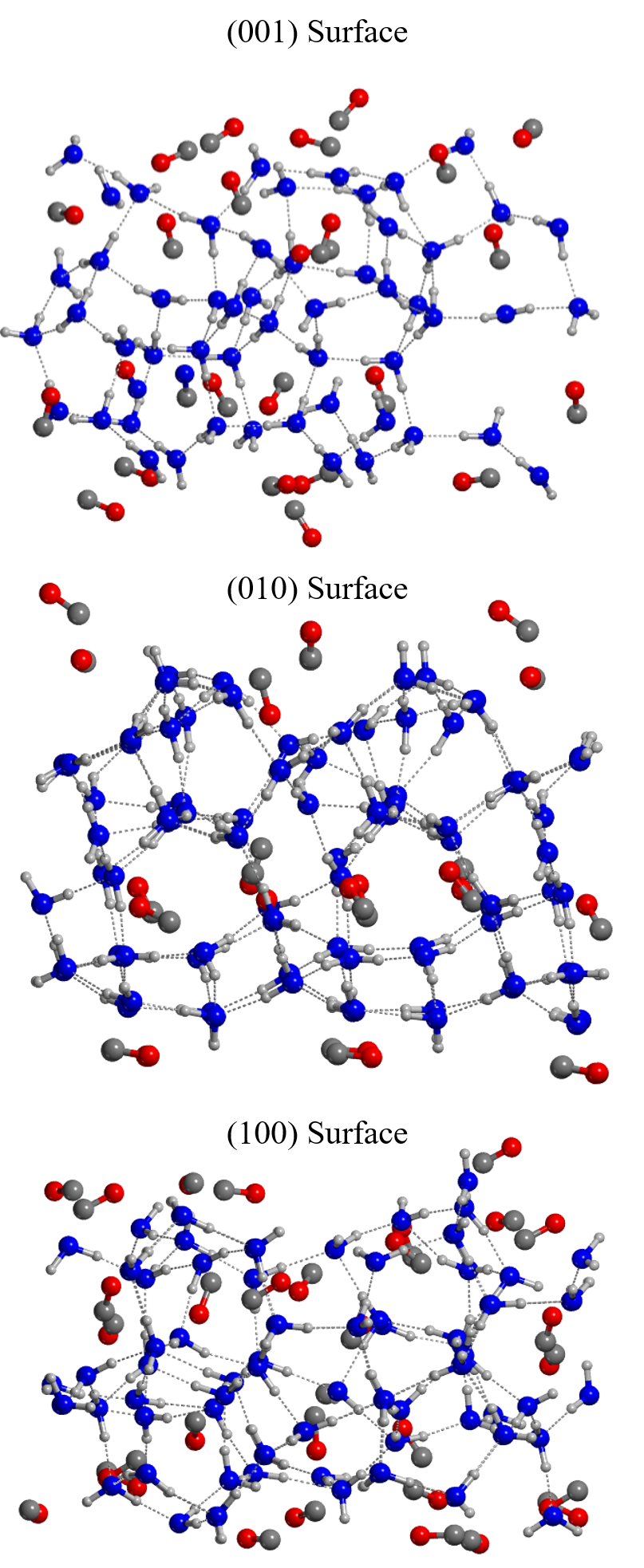}
    \caption{Structures of the (001), (010) and (100) ice surfaces, optimized at the HF-3c level of theory. Color code: H, white; C, grey; O atoms belonging to water, blue; O atoms belonging to the CO, red.}
    \label{fig:ice_surface}
\end{figure}

The H$_2$O:CO surfaces employed in this study were obtained from a bulk model of the pure H$_2$O crystalline P-ice \citep{casassa1997}, in which one every four water molecules was replaced by CO (Figure \ref{fig:bulk}). Three surfaces were cut along the planes (001), (010), (100) (see Figure \ref{fig:ice_surface}) and then fully relaxed at HF-3c level of theory, optimizing both cell parameters and atomic positions, which led to a heavy reorganization of each structure. 
Although we imposed a lattice to the system through periodic boundary conditions, we modeled large unit cells of H$_2$O:CO dirty ice (within P1 space symmetry) whose geometry relaxation resulted in an amorphous-like surface model.
The cell parameters, the dipole and the number of atoms of each structure can be found in Table \ref{tab:ice_surfaces}. The (001) surface has the smallest number of atoms (192) and the smallest dipole moment across the \textit{z} axis (-0.87 D at BHLYP-D3(BJ) level of theory). The H$_2$O and the CO molecules broke the symmetric structure of the bulk, creating clathrate-like cages in which CO molecules are surrounded by a network of water molecules engaged in hydrogen bonds (H-bonds). The outer layers of each surfaces are characterized by the presence of faintly interacting CO molecules that, depending on the coverage percentage, form a relatively complete monolayer of CO. Mixed ice models with a similar behaviour were reported in the theoretical work of \cite{zamirri2018}, in which CO adsorption, entrapment and mixture within H$_2$O ices was studied. There, it emerged that dispersive and quadrupolar forces have a prominent role in determining the structural features of these ice mixtures, and are responsible for the hydrophobic behaviour of CO. Indeed, when CO is entrapped in the ice structure, it causes large rearrangements of the network of hydrogen bonds, while when it is adsorbed on the top of the surfaces, it forms ordered layers. This gives rise to sections of the surface characterized by different electrostatic potential surfaces depending on the local arrangement of the carbon monoxide molecules.

\begin{table}
\centering
\caption{Cell parameters (a and b in \AA, and $\gamma$ in degrees) of the three ice surfaces optimized at HF-3c level. Dipole moment ($\mu$, in Debye) across the non-periodic \textit{z} axis of the surfaces calculated at HF-3c and BHLYP-D3(BJ) (DFT) level of theory. The number of water and carbon monoxide molecules of each surface per unit cell (H$_2$O:CO) is also reported.}
\label{tab:ice_surfaces}
\resizebox{\columnwidth}{!}{%
\begin{tabular}{lcccccc}
\hline
Surfaces &	a  &	b &	$\gamma$ &	$\mu$(HF-3c) &	$\mu$(DFT) & H$_2$O:CO \\
\hline
(100) &	14.718 &	12.718 &	101.294 &	-1.78 & -1.72 & 56:24 \\
(010) &	14.289 &	12.478 &	107.261 &	3.13 &	3.10 & 64:16 \\
(001) &	12.776 &	14.604 &	128.338 &	-1.52 &	-0.87 & 48:24 \\
\hline
\end{tabular}
}
\end{table}

In our model, the quantity of CO is not sufficient to give rise to a complete monolayer of CO, but the bottom face of (010) surface shows a neat arrangement of CO molecules. We notice very few cases in which CO molecules are engaged in short H-bonds with water molecules (< 2.4 \AA). This is because the water molecules tend to form a H-bond network between them and minimize the number of dangling H atoms pointing towards CO. Moreover, the majority of H-bond interactions take place through the O-end of the CO molecule and not through the C-end.

This phenomenon is probably driven by the small energy difference between the two interactions at HF-3c level of theory, for which the H-bond through C atom is favoured over the H-bond through the O atom by only 0.3 kJ mol$^{-1}$ (see Figure \ref{fig:dimer_h2o_co}). The C$\cdots$H interaction is also longer (2.390 \AA) than the O$\cdots$H one (2.212 \AA), although the results are comparable with those given by B3LYP-D3(BJ)/6-311G(d,p).

\section{Results}\label{sec:results}

\subsection{Preliminary benchmark studies}

\subsubsection{Gas-phase benchmark calculations}
In order to select the most appropriate DFT level of theory with which executing the periodic simulations, we firstly performed a benchmark analysis of the reactions \ref{chem_eqn:acetyl} and \ref{chem_eqn:acetaldehyde} in the gas-phase (namely, in the absence of the whole icy surfaces). For the sake of reliability of the benchmark study adopting these gas-phase reactions, geometry optimizations were performed at HF-3c level in CRYSTAL17, while the subsequent single point energies were computed with six different DFT methods, which were corrected with the Grimme's D3 or, when available, the D3(BJ) terms \citep{grimme:2010,grimme:2011} to account for dispersion interactions. Thus, the employed methods are: BHLYP-D3(BJ) \citep{bhandhlyp-becke1993,lee1988}, B3LYP-D3(BJ) \citep{lee1988,becke:1988,becke:1993}, M062X-D3 \citep{m062x-zhao}, M052X-D3 \citep{zhao:2006}, MPWB1K-D3(BJ) \citep{mpwb1k-zhao} and $\omega$B97X-D3 \citep{wb97x-chai}. They were used in conjunction with the Pople-based 6-311G(d,p) basis set. Diffuse functions were neglected because their small exponents can cause high linear dependencies in the wavefunction of the periodic systems we aim to work with. \citep{klanh1977,vandevondele2007,peintinger2013,oliveira2019}

\begin{figure}
    \centering
    \includegraphics[width=\columnwidth]{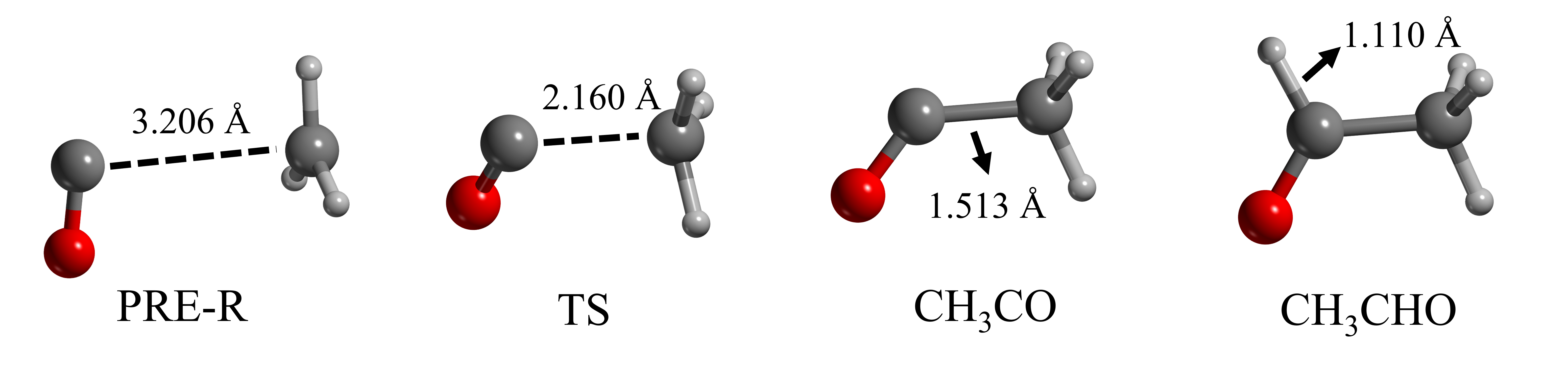} \\
    \caption{Stationary points of Reactions \ref{chem_eqn:acetyl} and \ref{chem_eqn:acetaldehyde}, computed at BHLYP-D3(BJ)/6-311G(d,p)//HF-3c level of theory. Distances are in \r{A}.}
    \label{fig:benchmark_pes}
\end{figure}

The DFT//HF-3c results were compared with those obtained by performing optimizations at the same DFT/6-311G(d,p) levels, followed by single point energy calculations at single- and double- electronic excitation coupled-cluster method with an added perturbative description of triple excitations (CCSD(T)) combined with Dunning's aug-cc-pVTZ basis set in Gaussian16 \citep{ccsdt}. The reaction path is represented in Figure \ref{fig:benchmark_pes}. Reaction \ref{chem_eqn:acetyl} presents a barrier due to being the coupling of 
the radical CH$_3$ with the closed-shell CO. Reaction \ref{chem_eqn:acetaldehyde} is a radical-radical coupling and, therefore, is barrierless in the gas-phase.

\begin{table*}
\centering
\caption{Relative energies (in kJ mol$^{-1}$) of the stationary points of the gas-phase reactions \ref{chem_eqn:acetyl} and \ref{chem_eqn:acetaldehyde} for all the DFT/6-311G(d,p) methods (computed with CRYSTAL17) and the CCSD(T)/aug-cc-pVTZ method (computed with Gaussian16). R stands for isolated reactants, PRE-R for pre-reactant complex, TS for transition state and P for product. The data are not ZPE-corrected.}
\label{tab:benchmark}
\resizebox{\textwidth}{!}{%
\begin{tabular}{|c|l|cccccc|c|}
\hline
Reaction            & Step           & A//H$^{\dagger}$  & B//H$^{\dagger}$  & C//H$^{\dagger}$ &  D//H$^{\dagger}$  & E//H$^{\dagger}$ & F//H$^{\dagger}$ & CCSD(T)//A$^{\dagger}$  \\
\hline 
%      BHLYP	B3LYP	M052X	M062X	MPWB1K	ωB97X   CCSD(T)
\multirow{7}{*}{R1: CO + CH$_3$ $\to$ CH$_3$CO} & R   &  0.03&  	-0.03&  	4.84&  	6.58	&  1.32	&  1.35 & 2.03  \\
                   & PRE-R    &  0.0 & 	0.0 & 	0.0	 & 0.0 & 	0.0 & 	0 & 0.0 \\
                   & TS      & 17.9 & 	0.2 & 	9.4 & 	15.6 & 	4.8	 & 3.5 & 22.5 \\
                    & P  &   -54.3 & 	-66.4	 & -58.0 & 	-51.1 & -74.9 & -71.6 & -56.1 \\
\cline{2-9}
& err\% TS R1        & 20.6\% &	99.3\% &	58.4\%	 &30.9\%	 &78.6\% &	84.3\%  &  \\
& err\% P R1 & 3.1\% & 18.5\% & 	3.4\% & 	8.9\% & 	33.5\%	 & 27.7\% & \\
\cline{2-9}
& mean err\% R1 & 11.9\% & 	58.9\% & 	30.9\% & 	19.9\% & 	56.1\% & 	56.0\% & \\

\hline 
\multirow{3}{*}{R2: CH$_3$CO + H $\to$ CH$_3$CHO} & R      & 0.0    & 0.0     & 0.0 & 0.0    & 0.0 & 0.0 & 0.0  \\
 & P  &  -390.7 &	-388.0 &	-394.8 &	-396.8 &	-383.0 &	-402.6 & -396.1 \\
\cline{2-9}
& err \% P R2   &  1.4\%	 & 2.0\% & 	0.3\% & 0.2\% & 3.3\% &	1.6\% &  \\
\hline
\end{tabular} %
}
\resizebox{\textwidth}{!}{
\begin{tabular}{l}
    $^{\dagger}$ A=BHLYP-D3(BJ), B = B3LYP-D3(BJ), C = M052X-D3, D = M062X-D3, E = MPWB1K-D3(BJ), F = $\omega$B97X-D3, H = HF-3c \\
\end{tabular}
}
\end{table*}

The results of this benchmark study for the gas-phase reaction models are shown in Table \ref{tab:benchmark}. The different CCSD(T)//DFT calculations return comparable energy values, meaning that results are almost unaffected by the small differences in the optimized geometries obtained with the different DFT methods. Thus, we only report here the results computed at CCSD(T)//BHLYP-D3(BJ), which can be taken as reference values. The error percentages on the energy barriers are high when comparing the DFT//HF-3c results with the CCSD(T)//BHLYP-D3(BJ) ones. To have deeper insights into that, we have analysed the resulting optimized structures with each method. Table \ref{tab:distances} reports the pivotal bond lengths of each species, where one can observe that HF-3c does not reproduce well the structures for the pre-reactant (PRE-R) and the transition state (TS) structures compared with the DFT methods. In the PRE-R and in the TS, the \ch{C\bond{single}C} length at HF-3c versus any DFT level shows differences going from 0.1 \AA~to 0.3 \AA~and, indeed, have a decisive impact on the energies that are computed afterwards at single point DFT.\\

\begin{table*}
\centering
\caption{Distances (in \AA) of selected bonds in the structures of the pre-reactants (PRE-R), transition state (TS) and products (CH$_3$CO and CH$_3$CHO) of the gas-phase reaction models \ref{chem_eqn:acetyl} and \ref{chem_eqn:acetaldehyde} computed at different theory levels.}

\label{tab:distances}
\resizebox{\textwidth}{!}{%
\begin{tabular}{|l|cccccc|c|}
\hline
Structure            &         BHLYP-D3(BJ) & B3LYP-D3(BJ) & M052X-D3 & M062X-D3  & MPWB1K-D3(BJ) & $\omega$B97X-D3  & HF-3c       \\
\hline
%      BHLYP	B3LYP	M052X	M062X	MPWB1K	ωB97X   HF-3c 
CO & 1.114 & 1.127 & 1.120 & 1.121 & 1.116 & 1.125 & 1.135 \\
\ch{C\bond{single}C} (PRE-R) & 3.369 & 3.190 & 3.211 & 3.206 & 3.431 & 3.294 & 3.154 \\
\ch{C\bond{single}C} (TS) & 2.146 & 2.238 & 2.171 & 2.147 & 2.243 & 2.209 & 2.015\\
\ch{C\bond{single}C} (CH$_3$CO) & 1.505 & 1.515 & 1.513 & 1.517 & 1.498 & 1.513 & 1.552 \\ % C-C
\ch{C\bond{single}O} (CH$_3$CO) & 1.166 & 1.180 & 1.174 & 1.173 & 1.167 & 1.177 & 1.187 \\ % C-O
\ch{C\bond{single}H} (CH$_3$CHO) & 1.513 & 1.114 & 1.106 & 1.110 & 1.105 & 1.113 & 1.107\\ % C-H
\ch{C\bond{single}O} (CH$_3$CHO) & 1.177 & 1.204 & 1.200 &  1.199 & 1.191 & 1.200 & 1.208 \\ % C-O
\hline
\end{tabular} %
}
\end{table*}

Energetic results indicate that B3LYP-D3(BJ) is the least suitable functional (99.3\% error on the potential energy barrier for Reaction \ref{chem_eqn:acetyl}), while BHLYP-D3(BJ) is the best one (20.6\% error though, corresponding to an absolute error of -4.6 kJ mol$^{-1}$). The BHLYP-D3(BJ) method also gives the smallest error percentage in the reaction energies for the formation of the CH$_3$CO radical. On the other hand, every functional describes well the reaction of CH$_3$CO with H.

\subsubsection{Grain-surface benchmark calculations}

Based on the error percentage on the energy of the gas-phase TS, we chose BHLYP-D3(BJ) to compute the reactions on the surface. However, although BHLYP-D3(BJ)//HF-3c performs globally well compared with CCSD(T)//BHLYP-D3(BJ), one has to pay special care when modeling the transition state of Reaction \ref{chem_eqn:acetyl}, which is the pivotal step to determine whether the reaction is feasible or not in the ISM. As probably the same discrepancies can affect  the periodic calculations, we also benchmarked them to assess the quality of the employed methods. Indeed, it is clear that we need to improve the quality of our data directly on the periodic surface, where the interaction between the reactants and the ice could change again the geometrical features of the interaction between CH$_3$ and CO, and therefore draw more discrepancies between HF-3c, BHLYP-D3(BJ) and CCSD(T) results.

A possible solution to this problem should be performing  ONIOM2 calculations \citep{dapprich:1999} combining BHLYP-D3(BJ) with CCSD(T), as low and high energy levels. This methodology has been previously applied to the computation of binding energies \citep{ferrero:2020,perrero2022be}. However, this would require obtaining optimized structures at BHLYP-D3(BJ) of the full periodic systems, which in our case is not feasible due to the high number of atoms present in the unit cells. Instead, we selected three test model cases onto which simulating the acetaldehyde formation and computing the energies applying the ONIOM2 refinement, in order to compare the performance of BHLYP-D3(BJ)//HF-3c against the ONIOM2-corrected values on the reaction barrier.

The test cases are based on two pure H$_2$O crystalline periodic ice models, in which one water molecule is replaced by a CO, and a molecular cluster of pure CO ice. They are shown in Figure \ref{fig:cbs_models} and are represented by: the 2x1 supercell of the (010) P-ice surface in which i) a water molecule exposing a dangling oxygen atom (\ch{H2O} Ice (a)) and ii) a water molecule exposing a dangling hydrogen atom (\ch{H2O} Ice (b)) was substituted by a CO molecule, and iii) a cluster model made of 20 CO molecules.

Each system was divided in two parts (\textit{model} and \textit{real} systems), described by two different levels of theory (\textit{high} and \textit{low}). The \textit{model} system (represented by the CH$_3$ and the CO) was described by the \textit{high} level of theory, CCSD(T). The \textit{real} system (that is, the whole system) was described by the \textit{low} level of theory, BHLYP-D3(BJ). In this ONIOM correction, CCSD(T) was used in combination with the Dunning's aug-cc-pVNZ (with N = D,T) basis sets \citep{dunning} and, with these data, the OAN(C) extrapolation scheme to the complete basis set (CBS) limit was applied \citep{okoshi2015}.

The ONIOM2-corrected energy barrier ($\Delta$E$_{TS}$) was computed as
\begin{multline} \label{ccsdt}
    \Delta E_{TS(ONIOM2)} = \Delta E_{TS}(low,real) + \\
    \Delta E_{TS}(high,model) - \Delta E_{TS}(low,model)
 \end{multline}

where the $\Delta$E$_{corr}$ = $\Delta$E$_{TS}$(\textit{high,model}) - $\Delta$E$_{TS}$(\textit{low,model}) represents the correction term to the energy of the \textit{real} system.

In this work, for the calculation of the ONIOM2-corrected barriers, $\Delta$E$_{ONIOM2}$, equation \ref{ccsdt} can be rewritten as:

\begin{multline}
    \Delta E_{TS(ONIOM2)} = \Delta E_{TS}(DFT;all) + \\
    \Delta E_{TS}(CCSD(T)/CBS;fragm) - \Delta E_{TS}(DFT;fragm) 
\end{multline}
 
where $\Delta$E$_{TS}$(DFT; \textit{all}) is the activation barrier computed at DFT//DFT. The $\Delta$E$_{TS}$ of the model system (CH$_3$ + CO; \textit{fragm}) is computed through single point energy calculations at CCSD(T)/aug-cc-pVNZ with n = D,T and extrapolated to the CBS limit thanks to the OAN(C) equation.

\begin{equation}
    E_{CBS}^{OAN(C)} = \frac{3^3E(T) - s^3E(D)}{3^3 - s^3}
\end{equation}

In this equation, s=2.091 based on the choice of method and basis set, E(T) is the energy calculated with the aug-cc-pVTZ basis set and E(D) corresponds to that computed with the aug-cc-pVDZ basis set.

We performed Reaction \ref{chem_eqn:acetyl} on the three test models, both at BHLYP-D3(BJ)//HF-3c level of theory and at the ONIOM2 scheme, chosen as reference. Data are available in Table \ref{tab:oniom2}.

\begin{table}
\centering
\caption{Potential energy barrier (in kJ mol$^{-1}$) of Reaction \ref{chem_eqn:acetyl} computed on the three test cases for the grain-surface benchmark. DFT stands for BHLYP-D3(BJ)/6-311G(d,p)}
\label{tab:oniom2}
\begin{tabular}{|l|cc|cc|c|}
\hline
 Structure       &   \multicolumn{2}{|c|}{\textit{real}}      & \multicolumn{2}{|c|}{\textit{model}}  & \textit{final} \\
        \hline
% Ice L becomes a; Ice R becomes b
 & DFT//HF-3c & DFT//DFT & DFT & CBS & ONIOM2 \\
H$_2$O Ice (a) & 27.0 & 49.5  & 13.8 & 19.6 & 55.2 \\ 
H$_2$O Ice (b) & 15.1 & 40.1 & 9.0 & 2.3 & 33.3\\ 
CO Ice & 11.3 & 15.0 & 18.0 & 20.4 & 17.4\\
\hline
\end{tabular}
\end{table}

\begin{figure}
    \centering
    \includegraphics[width=1.0\columnwidth]{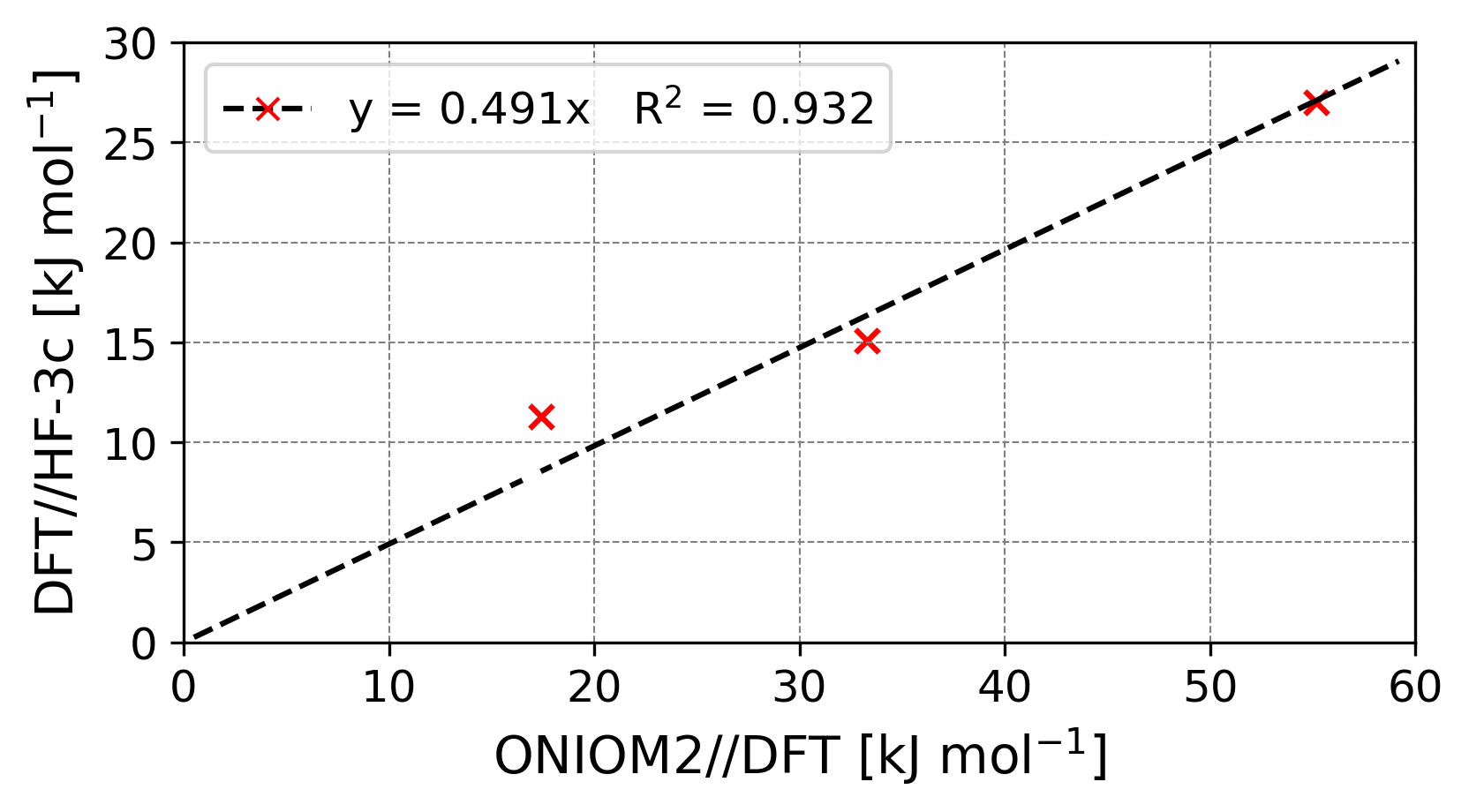}
    \caption{Plot of the DFT//HF-3c against ONIOM2//DFT potential energy barriers of Reaction \ref{chem_eqn:acetyl} performed on the three test models (the CO molecular cluster and the two crystalline periodic H$_2$O models with a CO substitution).}
    \label{fig:cbs_plot}
\end{figure}

In the plot of Figure \ref{fig:cbs_plot}, we compare the ONIOM2//DFT results against the DFT//HF-3c ones. The regression line (DFT//HF-3c=0.491$\cdot$ONIOM2//DFT, R$^2$=0.932) shows that the method of choice for the periodic calculations, DFT//HF-3c,  underestimates the reaction barrier, as  already emerging from the gas-phase benchmark study.

Since we have very few cases, we do not aim to adopt the slope as a correction factor. On the other hand, computing the reactions on the dirty ice surface models at full BHLYP-D3(BJ) level is almost unpractical. This benchmark study, however, allows to figure out the error bar associated with the present periodic calculations.

\subsection{Grain-surface reactions adopting an LH mechanism}

Once we have checked and chosen a reasonably suitable methodology to calculate Reactions \ref{chem_eqn:acetyl} and \ref{chem_eqn:acetaldehyde}, we simulated them on the dirty H$_2$O:CO icy surface models. 

To this end, we first adsorbed the CH$_3$ on the ice models by manually placing the radical in positions characterized by different local environments. Only the atomic positions were relaxed, while the cell parameters were kept frozen, to be consistent during the successive steps of the reaction and avoid structural deformations of the ice surface models. To calculate the binding energy (BE) of CH$_3$ on the mixed H$_2$O:CO surfaces, we followed the same computational scheme as in \cite{ferrero:2020} and \cite{perrero2022be}, that is, by correcting the adsorption energy $\Delta$E$_{ads}$ = E$_{complex}$ - E$_{ice}$ - E$_{CH_3}$ for the basis set superposition error (BSSE, which generates from using a finite basis set) through the counterpoise method by \cite{boysb:1970}. 

\begin{equation}
    BE = - \Delta E_{ads} + BSSE \label{eq:be}
    \end{equation}

For each case we computed the deformation energies of the surface and of the radical to verify that heavy structural rearrangements were not affecting our models during the geometry optimizations. 
We also computed the BE(0) at 0K, by correcting the BE for the $\Delta$ZPE as in equation \ref{eq:be_zpe}.

\begin{equation}
        BE(0) = BE - \Delta ZPE
     \label{eq:be_zpe}
\end{equation}

To identify the transition states, we adopted the distinguished reaction coordinate (DRC) procedure by performing a scan calculation along the \ch{C\bond{single}C} length. The maximum energy structure of the DRC pseudo-PES was used to localize and optimize the actual TS structure (as implemented in the CRYSTAL code, \citep{rimola2010}). In the DRC process, we avoided using internal redundant coordinates. Instead, we selected a mixed coordinate system made up of a set of selected valence internal parameters (bond lengths, angles and dihedrals) plus the full set of symmetry adapted fractional displacements and elastic distortions (evoked by the keyword INTLMIXED in CRYSTAL17). This represents an advantage both in reducing the number of valence internal parameters that are automatically generated for the system and in solving the quasi-linear dependencies that arises in systems with a lack of connectivity, which are reflected in a very high condition number of the Wilson B-Matrix.
The latter condition usually makes the optimization either to fail or to exhibit an erratic behavior \citep{crystal}.

To simulate Reaction \ref{chem_eqn:acetaldehyde}, we selected one newly formed CH$_3$CO/ice complex, and we manually placed the hydrogen atom to simulate its adsorption. We first set up five starting geometries and optimized the structures as  open-shell triplets (the two unpaired electrons with the same sign), which were then optimized as open-shell singlets (the two unpaired electrons with opposite signs). When the orientation of H was in favor of the formation of the \ch{C\bond{single}H} bond, but no acetaldehyde was obtained spontaneously, we performed DRC calculations to characterize the PES of the process.

\subsubsection{Adsorption of methyl radical on the dirty ices}

\begin{figure}
    \centering
    \includegraphics[width=1.0\columnwidth]{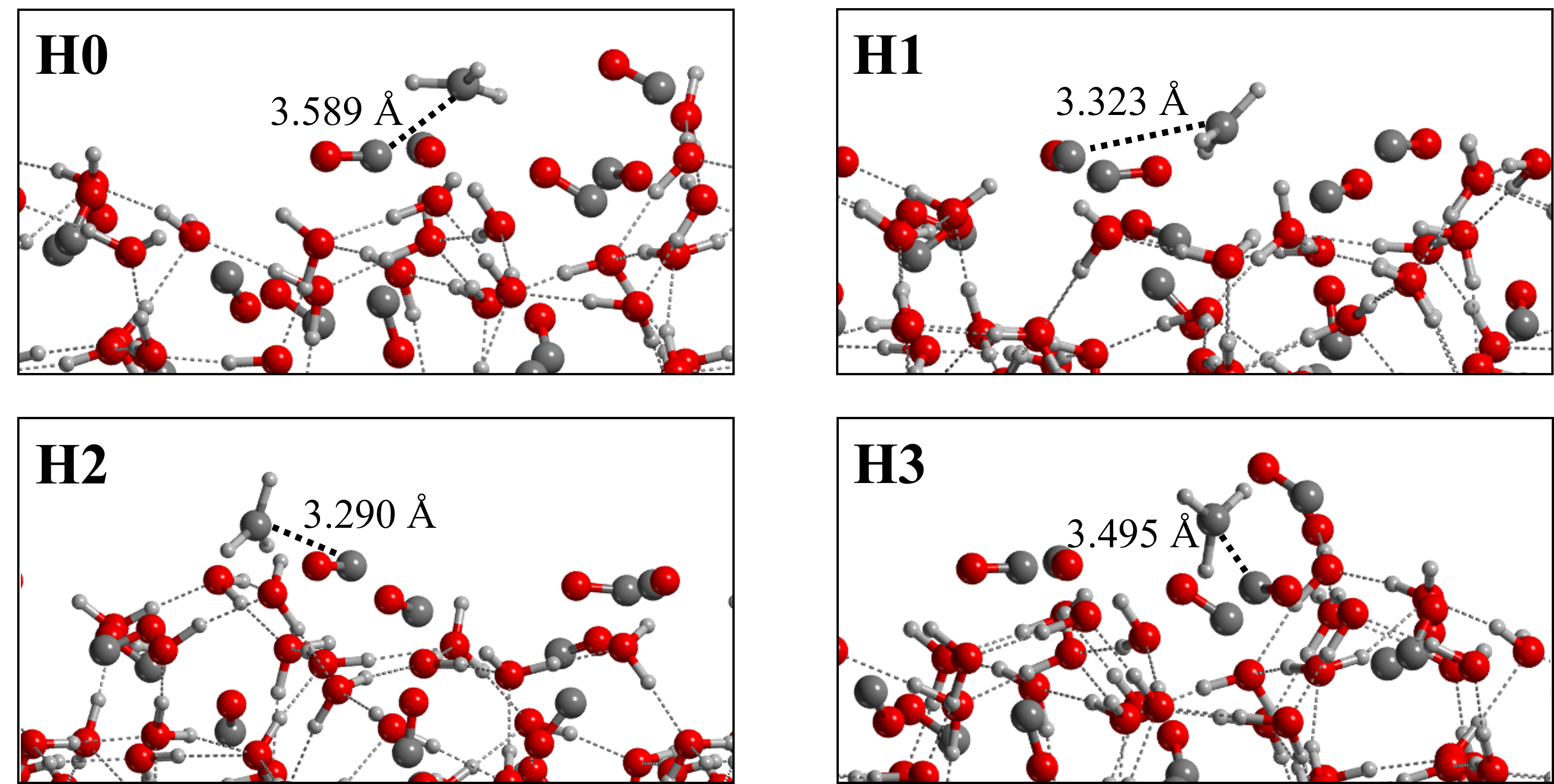}
    \caption{Structures of the \ch{CH3} adsorption complexes on the (100) surface, optimized at HF-3c level. Distances are in \AA.}
    \label{fig:adsorption}
\end{figure}

For each surface, CH$_3$ was adsorbed in four different positions (two on the top-face and two on the bottom-face) to sample different binding sites. The adsorption structures at the (100) surface are shown in Figure \ref{fig:adsorption}. 
Table \ref{tab:be} summarizes the computed BEs and their contribution, along with the adsorption enthalpy BE(0). The same Table \ref{tab:be} presents the nomenclature adopted for each binding site.

\begin{table*}
\centering
\caption{Binding energy (BE) values (in kJ mol$^{-1}$) of CH$_3$ on different surface sites of the dirty H$_2$O:CO ice model. The contributions from the pure potential energy values ($\Delta E_{ads}$), the BSSE corrections ($\Delta BSSE$), the dispersive (Disp) and the non-dispersive (No Disp) terms of the BE, the zero point energy corrections ($\Delta ZPE$) and the resulting adsorption enthalpy (BE(0)) are shown.}
\label{tab:be}
\begin{tabular}{ccccccccc}
\hline
Surface & Binding Site  & $\Delta E_{ads}$  & $\Delta BSSE$ & BE & Disp & No Disp & $\Delta ZPE$ &  BE(0)  \\
\hline  
\multirow{4}{*}{(100)}   & H0 &  -15.2 &    -5.9 &	9.3 & 15.0 & -5.7   &	4.0  &  5.3  \\
& H1 &  -18.8 & 	-8.9 &	9.8 & 17.8 & -8.0  & 	4.7 & 5.1 \\
& H2 &   -17.4 &   	-6.8 &	10.7 & 13.3 & -2.6   &	3.7 &  7.0 \\
& H3 &  -22.2 &  -10.8  &	11.3 & 20.2 & -8.8 &	4.7    &  6.7 \\
\hline  
\multirow{4}{*}{(010)} & K0  &  -27.3 &   	-13.4 &	14.0 &19.8&-5.8 &	 4.5  &   9.5\\
& K1 &  -14.5 & 	-6.2 &	8.3 &12.9&-4.6 & 	4.4 & 3.9\\
& K2 &   -31.8 &   	-13.4 &	18.4  &18.8&-0.4  &	4.7 &   13.7\\
& K3 &  -29.4 &    -11.0 &	18.5  &16.9&1.5  &	4.3 &  14.1  \\
\hline  
\multirow{4}{*}{(001)} & L0 &  -13.8 &   	-7.3 &	6.5 & 6.7&-0.2  &	5.5   & 1.0  \\
& L1 &  -13.0 &    -7.1 &	5.9  & 27.9&-22.0 &	6.6 & -0.7 \\
& L2 &  -18.9 & 	-7.6 &	11.3 &14.7&-3.3 & 3.9 & 7.4 \\
& L3 &   -40.3 &   	-6.9 &	33.4  &24.5&8.8  &	4.3 & 29.1   \\
\hline
\end{tabular}
\end{table*}

The methyl radical possesses an unpaired electron localized on the carbon atom, and its electrostatic potential surface is neutral almost everywhere. Therefore, this species will not form strong electrostatic interactions with the surface, especially with the polar water molecules. Thus, dispersion interactions are the key to explain the behavior of the methyl onto the surface. The BSSE can be as large as 50\% of the BE, due to the fact that the 6-311G(d,p) basis set has a rather small number of functions.

The BSSE-non-corrected adsorption energies vary between -13 and -40.3 kJ mol$^{-1}$. Considering the BSSE, the resulting BEs range from 5.9 to 33.4 kJ mol$^{-1}$. In nine out of the twelve characterized binding sites, the BEs are between 6 and 14 kJ mol$^{-1}$, while in two cases a value of 18.5 kJ mol$^{-1}$ is found. These values are similar to those obtained by \citet{ferrero:2020}, where the BE computed on the crystalline water ice is 18.2 kJ mol$^{-1}$, and on the amorphous water span the 9.2 to 13.8 kJ mol$^{-1}$ range, indicating a similar interaction of methyl radical with the two ice models. However, in the case of the binding site referred to as L3, the BE overcomes 30 kJ mol$^{-1}$. This is the only adsorption complex in which the interaction would be attractive even if we were not accounting for the dispersion. Indeed, while in all other cases the interaction would be repulsive, here the hydrogen atoms of the CH$_3$ are sufficiently close to the oxygen atoms of both CO and H$_2$O, therefore determining an advantageous electrostatic interaction of 8.8 kJ mol$^{-1}$. The ZPE correction is between 4-5 kJ mol$^{-1}$ in all the cases, as we would expect from the limited rearrangement of both the surface and the radical upon adsorption.

\subsubsection{Acetyl radical formation}

We modeled the formation of acetyl on a selected number of structures (see Figure \ref{fig:acetyl_formation} for an example). For each surface, we chose the three adsorption complexes characterized by the largest differences in the chemical environment of CH$_3$. The purpose of this choice is to probe the effect of the chemical environment on the potential energy barrier of the reaction. According to the LH mechanism, the CH$_3$ diffuses towards the closest CO on the surface to form the chemical bond. In all the nine cases analyzed (see Table \ref{tab:barriers}), a potential energy barrier has to be overcome in order to form the acetyl radical. The average barrier is around 10-15 kJ mol$^{-1}$, with some exceptions.
On the (100) surface, the complex \textbf{H3} has a barrier of only 5.8 kJ mol$^{-1}$. 
We notice that in this case, in the scan calculation, it is the CO that approaches the CH$_3$ and not vice versa, as we observed in the other simulations. 
We suppose that the particular geometry of this surface allows an easy diffusion of the CO, turning into a low potential energy barrier.
On the other hand, on the (010) surface, we found two unfavourable mechanisms, \textbf{K2} and \textbf{K3}, presenting energy barriers up to 36.8 kJ mol$^{-1}$.
In \textbf{K2}, a H-bond between the carbon of the CO and a water molecule of the surface needs to break to make CO available for the reaction with CH$_3$. In \textbf{K3}, the large distance (3.9 \AA) between the reactants is the responsible of the high barrier.

When adding the ZPE contributions, according to the equation $\Delta$H$_{TS}$ = $\Delta$E$_{TS}$ + $\Delta$ZPE, each barrier increases by about 11 kJ mol$^{-1}$. For the gas-phase reaction, we found barriers of $\Delta$E$_{TS}$ = 17.9 kJ mol$^{-1}$ and $\Delta$H$_{TS}$ = 28.7 kJ mol$^{-1}$, meaning a $\Delta$ZPE$_{TS}$ = 10.8 kJ mol$^{-1}$, very close to that obtained on the surface.
If we focus solely on the energy barrier, in seven cases out of nine the barriers of the surface reactions are lower than the gas-phase one. However, the grain-surface benchmark calculations warn us that these barriers are underestimated and, therefore, it is highly probable that only the grain-surface \textbf{H3} case is slightly more favourable than the gas-phase reaction.

\begin{table}
\centering
\caption{ZPE-corrected potential energy barriers ($\Delta$H(0)$_{TS}$, in kJ mol$^{-1}$) for the formation of acetyl on different binding sites of each surface. The internal potential energy values ($\Delta E_{TS}$) and the zero point energy corrections ($\Delta ZPE_{TS}$) are displayed.}
\label{tab:barriers}

\begin{tabular}{ccccc}
\hline
Surface & Binding Site  & $\Delta E_{TS}$  & $\Delta ZPE_{TS}$ & $\Delta H(0)_{TS}$  \\

\hline  
\multirow{3}{*}{(100)}  & H1 &  11.7 & 10.7 & 22.4 \\
& H2 &   13.4 & 11.3 &  24.7  \\
& H3 &  5.8 & 11.0 &  16.8  \\
\hline  
\multirow{3}{*}{(010)}  & K1 &  11.6 & 10.3   & 21.9\\
& K2 &   23.7 & 10.0 & 33.7\\
& K3 &  36.8 & 10.1 & 46.9 \\
\hline  
\multirow{3}{*}{(001)}  & L1 &  10.8 & 10.8 & 21.6  \\
 & L2 &  14.5 & 12.3 & 26.8 \\
 & L3 &   14.4 & 11.9  & 26.5\\
\hline
\hline
Gas-phase & - & 17.9 & 10.8 & 28.7 \\
\hline
\end{tabular}

\end{table}

\begin{figure}
    \centering
    \includegraphics[width=0.95\columnwidth]{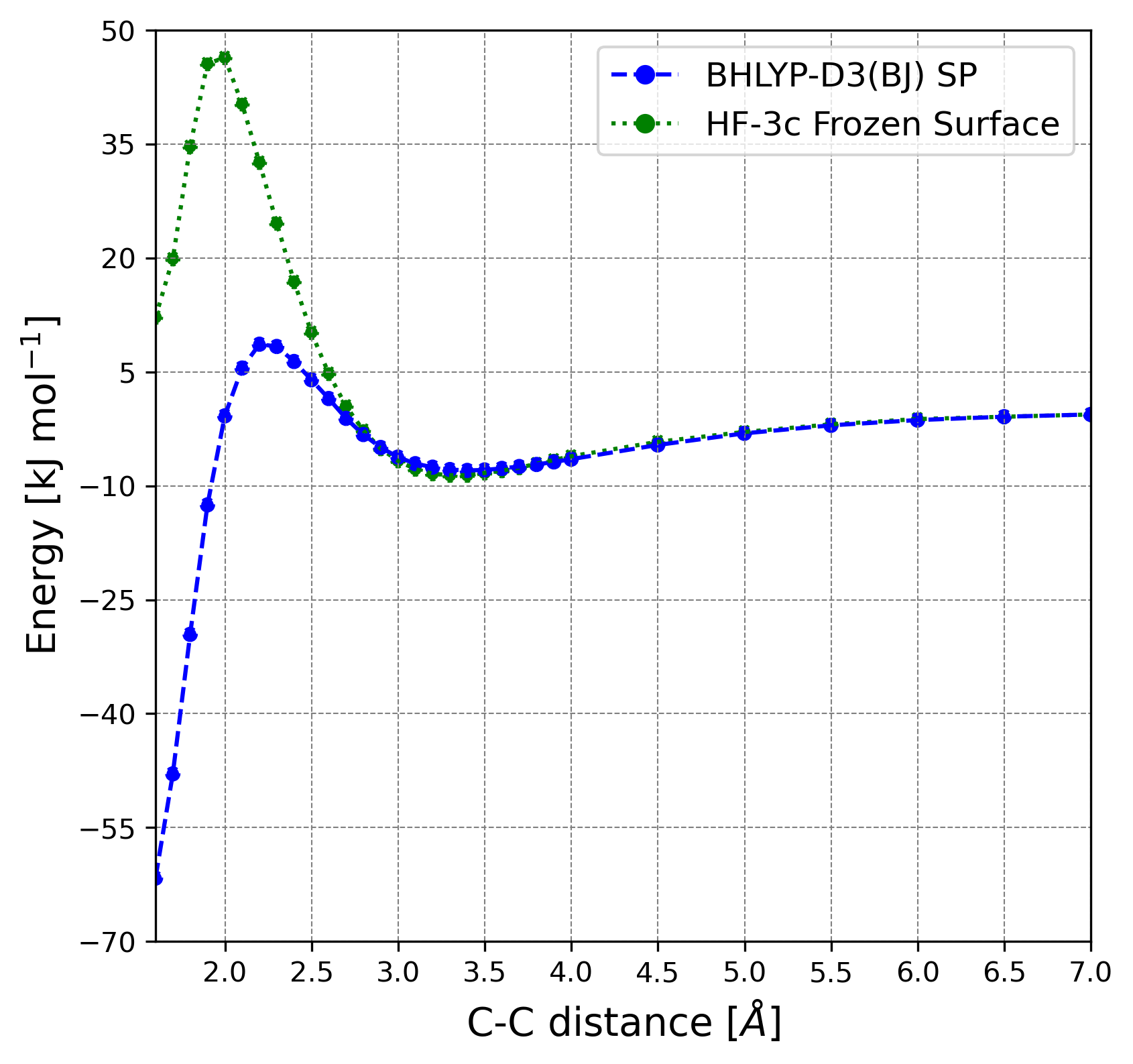}
    \caption{Eley-Rideal reaction profile of CH$_3$ + CO. Energy is given in kJ mol$^{-1}$, \ch{C\bond{single}C} distance in \AA. The reaction is computed keeping the surface frozen at HF-3c level of theory (green). We also provide the single point energy computed at BHLYP-D3(BJ)/6-311G(d,p) level of theory (blue) on HF-3c geometries.}
    \label{fig:ER_I}
\end{figure}
\begin{figure}
    \centering
    \includegraphics[width=0.95\columnwidth]{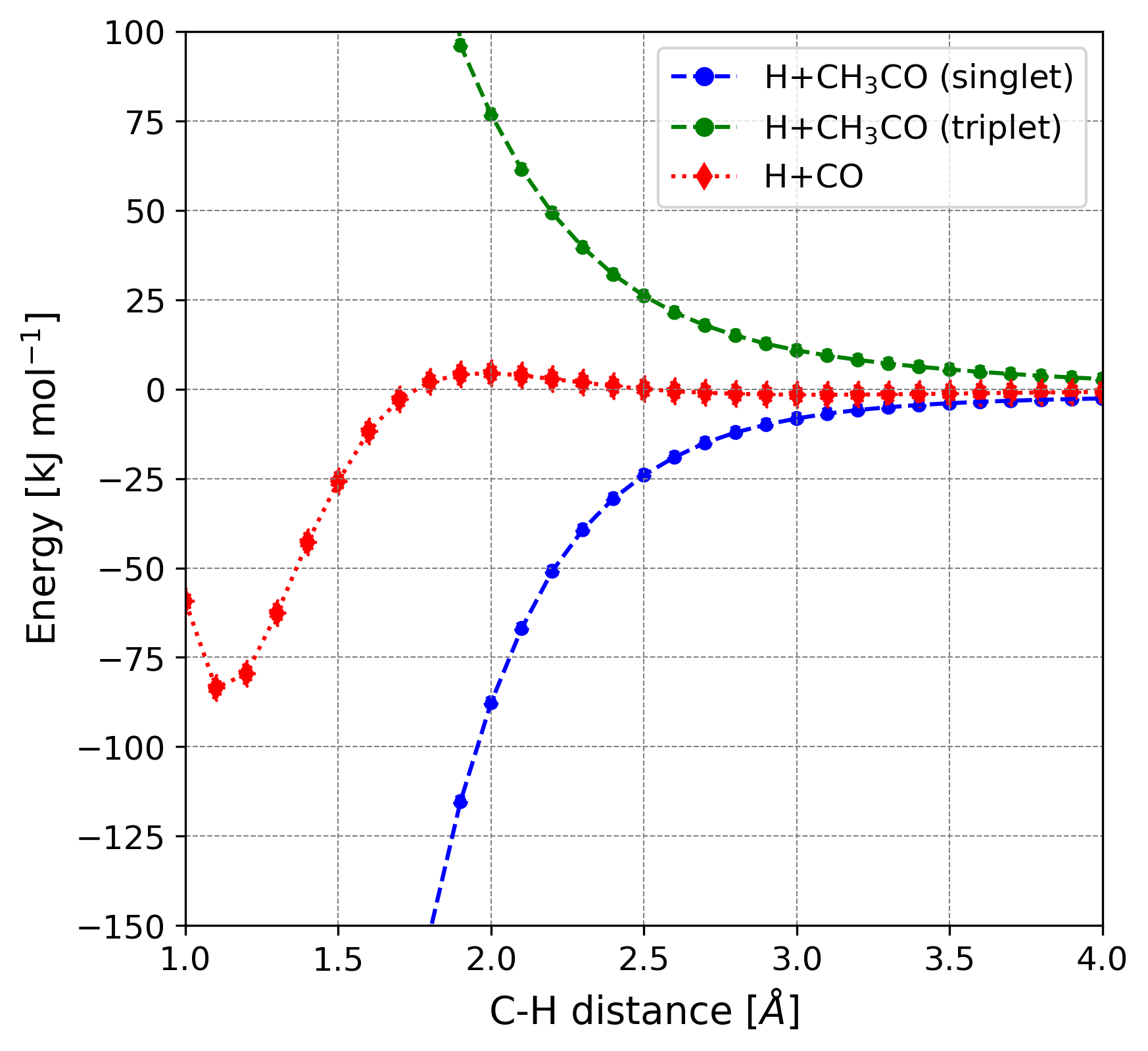}
    \caption{Eley-Rideal reaction profile of H + CH$_3$CO and its competitive reaction H + CO at BHLYP-D3(BJ)/6-311G(d,p) level of theory. Energy is given in kJ mol$^{-1}$, \ch{C\bond{single}H} distance in \AA. H + CO presents a small barrier (red), H + CH$_3$CO in the triplet spin state (green) becomes repulsive, while in the singlet spin state (blue) yields the product barrierlessly.}
    \label{fig:ER_II}
\end{figure}

\subsubsection{Acetaldehyde formation}
Among the different acetyl products obtained, we selected the \textbf{H1} case to model Reaction \ref{chem_eqn:acetaldehyde}. The reason of this choice is because, in this structure, the acetyl is bound to a water molecule through a H-bond involving the oxygen of the carbonyl group, namely, the carbon atom of interest is not hindered because it is pointing towards the gas-phase (it is not facing the inner side of the surface), and thus it is prone to react with a hydrogen atom diffusing on the surface (see Figure \ref{fig:acetaldehyde_formation}). 

The adsorption of atomic hydrogen followed by the optimization of the CH$_3$CO + H complex in an open-shell triplet spin state resulted, as expected, in no reaction. However, a change of the electronic spin state to an open-shell singlet brought to the spontaneous formation of acetaldehyde in two (out of five) complexes without harming the H-bond interaction between the carbonyl moiety and the H$_2$O$_{(ice)}$. We noticed that if the H atom is at a maximum distance of 3.5 \AA~from the CO moiety and it is correctly oriented (meaning that it approaches the CO from the less hindered side), the radical coupling is barrierless. 

In the other cases, either the H atom is hindered by the methyl moiety, or it has to overcome a small diffusion barrier to get close enough to the reactive center. With the methodology chosen for the optimization, we notice that a very small diffusion barrier, of about 1 kJ mol$^{-1}$, is limiting the free diffusion of the hydrogen on the surface, as the gradient in the optimization process falls to zero and a minimum is found.

\subsection{Grain-surface reactions adopting an ER mechanism}

In the ER mechanism, the CH$_3$ gas-phase species directly reacts with a surface CO icy component. To simulate this, we generated, with a Python script, several geometries in which the distance between the approaching CH$_3$  from the gas-phase and the reactive CO of the surface progressively decreases. We ran the optimization of these structures without relaxing the geometry of the surface to avoid its thermalization during the approach of the CH$_3$. Only after the product forms, the entire system was relaxed. Therefore, we provide the energy profiles for Reactions \ref{chem_eqn:acetyl} and \ref{chem_eqn:acetaldehyde} without characterizing the actual transition states but only giving an estimate. This is because, due to the geometrical constraints applied to the surface, we would obtain high order stationary points with imaginary frequencies associated with the motion of the frozen atoms of the surface. For the same reason, we also do not provide the ZPE-corrected profiles for these processes. 
We simulated both steps of the reaction assuming an ER mechanism. We chose the (100) surface as the reactive one, focusing on the \textbf{H1} binding site. We did so in order to compare LH and ER mechanisms for steps \ref{chem_eqn:acetyl} and \ref{chem_eqn:acetaldehyde}, but also because the CO involved in the \ch{C\bond{single}C} bond formation is exposing the C-end towards the gas-phase, therefore being available for this type of reactivity.
The energy profile of CH$_3$ + CO is plotted in Figure \ref{fig:ER_I}. Both the profile computed at HF-3c level and the one refined at BHLYP-D3(BJ) level are shown. One can clearly see the disagreement between the two methodologies in terms of energetics. The BHLYP-D3(BJ) profile shows that below 5 \AA~of distance there is an attractive interaction between the methyl radical and the CO. The presence of a van der Waals complex well in the energy profile appears at a distance of 3.3 \AA. When getting close to 2 \AA, we find the maximum energy point, whose value is about 8.5 kJ mol$^{-1}$ compared with the energy at infinite distance. The $\Delta$E between the minimum and maximum energy point is 16 kJ mol$^{-1}$, which is compatible with the barrier computed for the LH mechanism. According to these results, with both the formation of such complex and the presence of a barrier, the reaction will hardly yield acetyl.  

On the other hand, considering that acetyl is already on the surface and a hydrogen atom approaches from the gas-phase, two situations can take place. If the overall spin of the system is a triplet, no coupling takes place and the energy of the system becomes highly repulsive as the distance between the two reactants increase. In contrast, if the system is in a singlet open-shell electronic state, the formation of the \ch{C\bond{single}H} bond occurs spontaneously. 
We also simulated the competitive reaction, in which H falls onto the surface and reacts with an adjacent surface CO molecule (considering an overall triplet spin state) forming the HCO radical with a small barrier of 4.5 kJ mol$^{-1}$. Thus, in case of a H atom approaching the surface, the formation of acetaldehyde via H addition to acetyl is the most probable reaction. 

%%%%%%%%%%%%%%%%%%%%%%%%%%%%%%%%%%%%%%%%%%%%%%%%%%%%%%%%%%%%%%%%
\section{Discussion and Astrophysical Implications}\label{sec:discuss}

In this work, we simulated the formation of acetaldehyde on H$_2$O:CO dirty ice surfaces models through the reaction of a methyl radical with a CO belonging to the ice surface, in which, subsequently, the newly formed acetyl radical gets hydrogenated to yield acetaldehyde. We performed these two-steps process on the basis of both LH and ER surface mechanisms.  

Thanks to this "radical + ice" reaction mechanism, step \ref{chem_eqn:acetyl} has no competitive reactions, at variance with the prevailing radical-radical coupling mechanism (\citet{enrique-romero2022}). Indeed, in our case, CH$_3$ cannot extract a hydrogen atom from water to yield CH$_4$, neither we consider such a high abundance for CH$_3$ to be able to form ethane (CH$_3$CH$_3$). 
As far as Reaction \ref{chem_eqn:acetaldehyde} is concerned, there may be a competitive channel, i.e., H + CO $\rightarrow$ HCO, known to have a barrier of $\sim5$ kJ mol$^{-1}$ and to occur through tunnelling \citep[e.g.][]{rimola2014, pantaleone2020}. 
In our simulations, we also observe the presence of a barrier. Even though we modeled the process through an ER mechanism and we did not isolate the exact transition state of this reaction, the energy profile reaches its maximum of energy at 4.5 kJ mol$^{-1}$. In contrast, the formation of acetaldehyde is barrierless and, therefore, favoured. 

The benchmark that we performed for the acetyl formation reaction on the icy surfaces stresses that the BHLYP-D3(BJ)//HF-3c computational scheme seems to underestimate the potential energy barrier of the process. 
Thus, a direct comparison with the gas-phase reaction (computed at full DFT level) is, therefore, not possible. 
However, if we roughly assume that the computed $\Delta$E(0)$_{TS}$ values on the surfaces have to be doubled to be fairly compared with the gas-phase ones (as emerged from the benchmark study), it turns out that the grain-surface reactions are less favourable than the gas-phase ones in the majority of the cases, namely, just one case out of nine is characterized by a $\Delta$H(0)$_{TS}$ lower than the 28.7 kJ mol$^{-1}$ found for the gas-phase reaction. 
However, even in the most favourable case, the barrier is still high so that Reaction \ref{chem_eqn:acetyl} is unlikely to take place in the cold molecular clouds. 
Additionally, in environments where the temperature can be higher (namely, with more thermal energy available to overcome the barrier), the sublimation of CO and CH$_3$ \citep[above 20 K: e.g.][]{ferrero:2020} will prevent the reaction from taking place.

Therefore, we can summarize the effect of the ice surface on the acetyl formation through the reaction of CH$_3$ with iced CO in two points: i) the CO embedded in the surface is not strongly activated towards the reaction by the surrounding icy water molecules, which in fact, if possible, avoid interacting with the CO molecules by creating clathrate-like structures, as it appears during the optimization of the surfaces; and ii) in almost all the surface reactions, the potential energy barrier is fairly larger than that in the gas-phase, meaning that there is an additional, although modest, interaction between the CH$_3$ radical and the surface that needs to be broken in order to allow the radical to approach the CO and form a \ch{C\bond{single}C} bond. 

On the other hand, the second step (formation of acetaldehyde by H-addition to acetyl) is favoured by the presence of the ice, as the surface is well known to act as a reactant concentrator \citep[e.g.][]{ioppolo2011,rimola2014, Fedoseev2015,Simons2020,ferrero2023}
as well as third body (i.e., energy dissipator, hence stabilising the newly formed product) \citep[e.g.][]{pantaleone2020, pantaleone2021,ferrero2023nh3,molpeceres2023} in hydrogenation of atoms and small molecules. Moreover, in the event that hydrogen is approaching the acetyl with an unfavourable orientation, thanks to its easy diffusion, it is able to move towards the CO moiety and yield acetaldehyde.

The ER mechanism does not bring any improvement over the LH one. For the former, to be effective, the acetyl formation should not have a barrier, this way avoiding the preliminary adsorption and forming directly the product. \cite{Ruaud2015} stresses the importance of this mechanism for reactions between atomic carbon and icy components. However, in our case, both mechanisms present a barrier with similar heights, so that none of the mechanisms dominate over the other (at least from an energetic standpoint). For the hydrogenation of the acetyl radical to form acetaldehyde, both mechanism are feasible to yield the product, although the LH one is probably favoured due to the fact that at a temperature as low as 10 K the majority of atomic hydrogen would probably be adsorbed on the surface. We have seen, moreover, that the relative orientation between H and acetyl is not a hampering factor so that probably, at 10 K, H atoms have enough mobility to jump between different adsorption sites and find the orientation that would favour the formation of acetaldehyde.

Therefore, although the CH$_3$ + CO$_{(ice)}$ mechanism seems not to particularly enhance the formation of acetaldehyde, this result is important to constraint the active synthetic paths that drive the formation of acetaldehyde in the ISM. 

Other studies have focused on acetaldehyde grain-surface formation through radical recombination between HCO and CH$_3$ \cite[e.g.][]{enrique-romero2016, enrique-romero2019, enrique-romero2021, Lamberts2019}, which already highlighted the difficulty of the synthesis of this iCOM.
In \cite{enrique-romero2016} and \cite{enrique-romero2019}, the reactivity between CH$_3$ and HCO performed on H$_2$O ice resulted in either formation of acetaldehyde or in CH$_4$ + CO due to a competitive H abstraction, both reactions having small or no potential energy barriers. 
In a subsequent work, \cite{enrique-romero2021} found that the efficiency of acetaldehyde formation was overall low in comparison with that of CH$_4$ + CO. 
In \cite{Lamberts2019}, the radical pair reactivity was investigated on CO ices by means of \textit{ab initio} molecular dynamics. 
By adopting a sufficient configurational sampling (namely, different trajectories based on different initial guess structures), the authors found that the reactivity results in either no reaction, formation of CH$_3$CHO or CH$_4$ + CO, the last two outcomes being barrierless, while in the first case the non-formation of a product was due to the presence of a barrier. 
Therefore, according to these results, acetaldehyde synthesis on icy surfaces is not a favourable path. 
This is in agreement with the theoretical study by \cite{Simons2020}, who stated that the reaction network obtained by the hydrogenation of CO may cause the production of several iCOMs (glycolaldehyde, ethylene glycol and to a less extent methyl formate), alongside methanol and formaldehyde, but acetaldehyde is not one of them.
Likewise, the experimental study by \cite{gutierrez2021} found that acetaldehyde is not formed by the combination of radicals created by the UV illumination of methanol ice.

An alternative way of forming acetaldehyde on the grain-surfaces discussed in the literature is via the reaction of carbon atoms with the CO molecules of the ices.
\cite{Fedoseev:2022} experimentally studied the reaction of C + CO co-deposited with H$_2$O and H at 10 K, following the theoretical study of \cite{papakondilis:2019}. 
The authors observed the formation of CCO and its hydrogenated counterpart, the ketene CH$_2$CO. 
Successive hydrogenation steps result in CH$_3$CHO. 
Likewise, \cite{ferrero2023} carried out a theoretical study on the formation of ketene from the reaction of C with CO$_{(ice)}$ and the potential successive reactions with radicals (e.g. OH and NH$_2$) that could form other iCOMs and found that only hydrogenation (via H-tunneling), eventually leading to acetaldehyde could occur, because of important energy barriers.
However, as \cite{ferrero2023} noticed, the simultaneous presence of gaseous atomic carbon landing on the grain-surfaces and a CO-rich ice is very unlikely in the molecular ISM, except in Photo-Dissociation Regions (PDRs), as abundant frozen CO implies an evolved molecular cloud or prestellar core where atomic carbon has an extremely low abundance.

Assuming that CH$_3$CHO is formed on the grain-surfaces, then a thermal and/or non-thermal mechanism is needed to partially transfer it into the gas-phase, where it is observed.
This is particularly problematic when considering the detection of acetaldehyde in several cold prestellar cores \citep{Scibelli2020}.
The parameter establishing whether a species stays bound to the icy mantle or enriches the gas-phase is its BE. 
A recent computational work by \cite{ferrero2022} highlights how the acetaldehyde BE, ranging from 3000 K to 7000 K on water ice, represent an obstacle to explain its presence in the gas-phase of cold environments. 
The studies by \cite{corazzi2021} and \cite{molpeceres2022} reported a value which is close to the lower end of the distribution outlined in \cite{ferrero2022}, but it is still too large to allow desorption in cold (10 K) astronomical objects. 
Usually, to explain the desorption process of this and other iCOMs, non-thermal mechanisms such as photo-desorption, reactive desorption, and cosmic-ray desorption are invoked \cite[e.g.][]{dulieu2013,chuang2018,dartois:2019}, each of these mechanism having their drawbacks \citep[e.g.][]{Bertin2016-photodes,pantaleone2020}. 
Clearing this matter is out of the scope of this work, which aims at investigating acetaldehyde formation on icy surfaces.

In summary, all studies so far carried out, theoretical and experimental, tend to agree that acetaldehyde is unlikely to be a grain-surface product.
On the contrary, several studies now seem to favor the hypothesis of acetaldehyde formed in the gas phase.
We would like to mention in particular the work by \cite{vazart2020}, who showed a very good agreement between the observed and measured abundance of acetaldehyde in hot corinos.

%%%%%%%%%%%%%%%%%%%%%%%%%%%%%%%%%%%%%%%%%%%%%%
\section{Conclusions} \label{sec:conclusions}

In this work, we studied the formation of acetaldehyde (CH$_3$CHO) on the grain surfaces through a two-steps mechanism consisting of: i) the reaction of a methyl (CH$_3$) radical with a CO molecule belonging to a periodic surface of H$_2$O:CO dirty ice to form acetyl (CH$_3$CO), and ii) the hydrogenation of the newly formed acetyl. 
We characterized the process via the Langmuir-Hinshelwood and Eley-Rideal mechanisms and compared the results obtained against those simulated for the same reaction in the gas-phase. 
Our grain-surface benchmark on three test models stressed that the barriers computed at BHLYP-D3(BJ)//HF-3c level are fairly underestimated against CCSD(T)//BHLYP-D3(BJ). 

We modeled three periodic dirty ice surfaces by cutting along the planes (001), (010), and (100) a bulk of crystalline H$_2$O P-ice, where one fourth of the water molecules was substituted by CO. 
We obtained three amorphous-like surfaces characterized by a variable content in CO onto which we simulated several adsorption structures of CH$_3$, finding that dispersion interactions are crucial for the adsorption process. 
We computed the ZPE-corrected potential energy surfaces of acetyl formation in three cases for each H$_2$O:CO ice surface, results pointing out that in most of the cases the reaction presents high energy barriers insurmountable in the ISM. 
The formation of acetyl via CH$_3$ + CO$_{(ice)}$ on the surface is less favourable than its gas-phase counterpart in the majority of the cases. 

We have elucidated that these overall unfavourable reactions are due to that: i) the CO molecule is hardly activated by the surface and the CH$_3$ does not present an enhanced reactivity (it does not form hemibonded systems), and ii) the barriers for acetyl formation depend on the local environment of the CO, that is, they are high (in most of the cases) when the distance between the reactants is large or when the CO/H$_2$O interactions have to be broken for the reaction to take place. 
On the other hand, the hydrogenation of acetyl is barrierless, given that the electronic spin of the system is a singlet open-shell and that the H atom is so mobile that it always finds a proper orientation towards the carbonyl group of CH$_3$CO. 
The outcome of the reaction, specifically its barrier, does not improve when adopting the ER mechanism against the LH one.

In summary, several previous studies have challenged the hypothesis that acetaldehyde is formed on the grain-surfaces via the combination of on-surface radicals, specifically CH$_3$ and HCO \cite[e.g.][]{enrique-romero2016, enrique-romero2019, Lamberts2019, Simons2020, enrique-romero2021, gutierrez2021}.
Leveraging on other studies showing that some iCOMs \citep[ethanol and formamide:][]{perrero2022ethanol,Rimola2018} could be formed on the grain surfaces via reactions of radicals with molecules belonging to the ice, in this work, we investigated the reaction of the CH$_3$ radical with one CO molecule of the ice.
Our computations show that also this path is unfavorable to the acetaldehyde formation, reducing the possibilities that the latter is formed on the grain surfaces.
Alternatively, gas-phase reactions could be at the origin of the almost ubiquitous presence of acetaldehyde in the ISM \citep[e.g.][]{vazart2020}.

\section*{Acknowledgements}

This project has received funding within the European Union’s Horizon 2020 research and innovation programme from the European Research Council (ERC) for the projects ``Quantum Chemistry on Interstellar Grains” (QUANTUMGRAIN), grant agreement No 865657 and ``The Dawn of Organic Chemistry” (DOC), grant agreement No 741002. The authors acknowledge funding from
the European Union’s Horizon 2020 research and innovation program Marie Sklodowska-Curie for the project ``Astro-Chemical Origins” (ACO), grant agreement No 811312. MICINN (project PID2021-126427NB-I00) and Italian MUR (PRIN 2020, Astrochemistry beyond the second period elements, Prot. 2020AFB3FX) are also acknowledged. CSUC supercomputing center is acknowledged for allowance of computer resources.
We thank Prof. Gretobape for fruitful and stimulating discussions. J.P. is indebted to Joan Enrique-Romero for his help in python scripting. 

%%%%%%%%%%%%%%%%%%%%%%%%%%%%%%%%%%%%%%%%%%%%%%%%%%
\section*{Data Availability}

The data underlying this article are freely available in Zenodo
at \url{https://doi.org/10.5281/zenodo.7937759}.

%%%%%%%%%%%%%%%%%%%% REFERENCES %%%%%%%%%%%%%%%%%%

% The best way to enter references is to use BibTeX:

\bibliographystyle{mnras}
\bibliography{MY_BIBLIO}

%%%%%%%%%%%%%%%%%%%%%%%%%%%%%%%%%%%%%%%%%%%%%%%%%%

%%%%%%%%%%%%%%%%% APPENDICES %%%%%%%%%%%%%%%%%%%%%
\clearpage
\appendix

\section{Benchmark models}

\begin{figure}
    \centering
    \includegraphics[width=0.6\columnwidth]{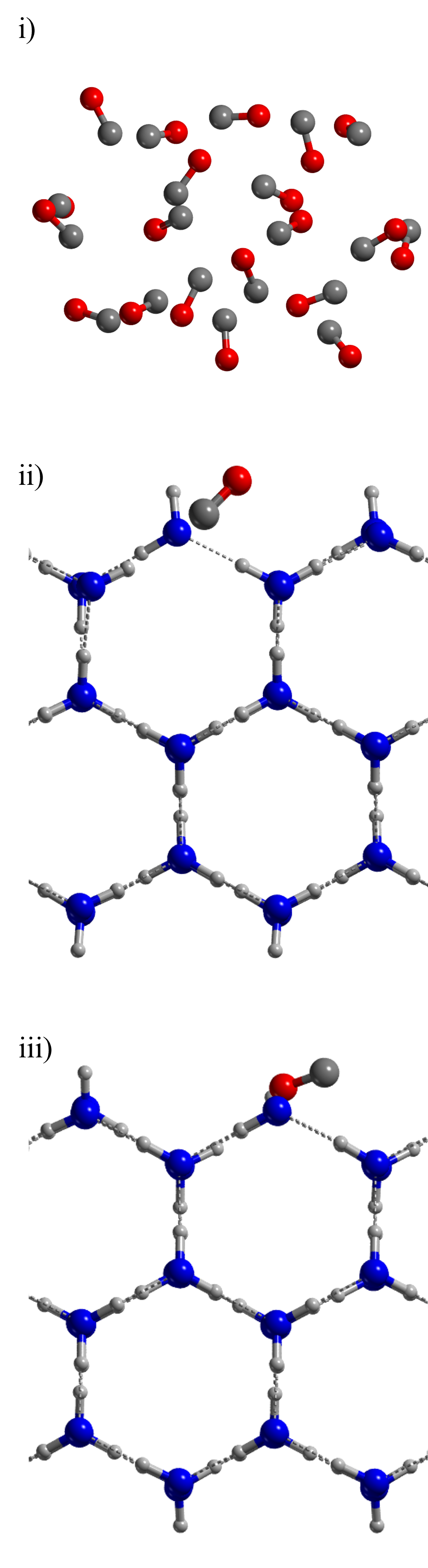}
    \caption{Ice model used for the benchmark on the surface: i) a cluster model made of 20 CO molecules, and the 2x1 supercell of (010) P-ice surface in which ii) a water molecule exposing a dangling oxygen atom (\ch{H2O} Ice (a)) and iii) a water molecule exposing a dangling hydrogen atom (\ch{H2O} Ice (b)) was substituted by a CO molecule.}
    \label{fig:cbs_models}
\end{figure}

\clearpage

\section{\ch{H2O} $\cdots$ CO interaction}

\begin{figure}
    \centering
    \includegraphics[width=\columnwidth]{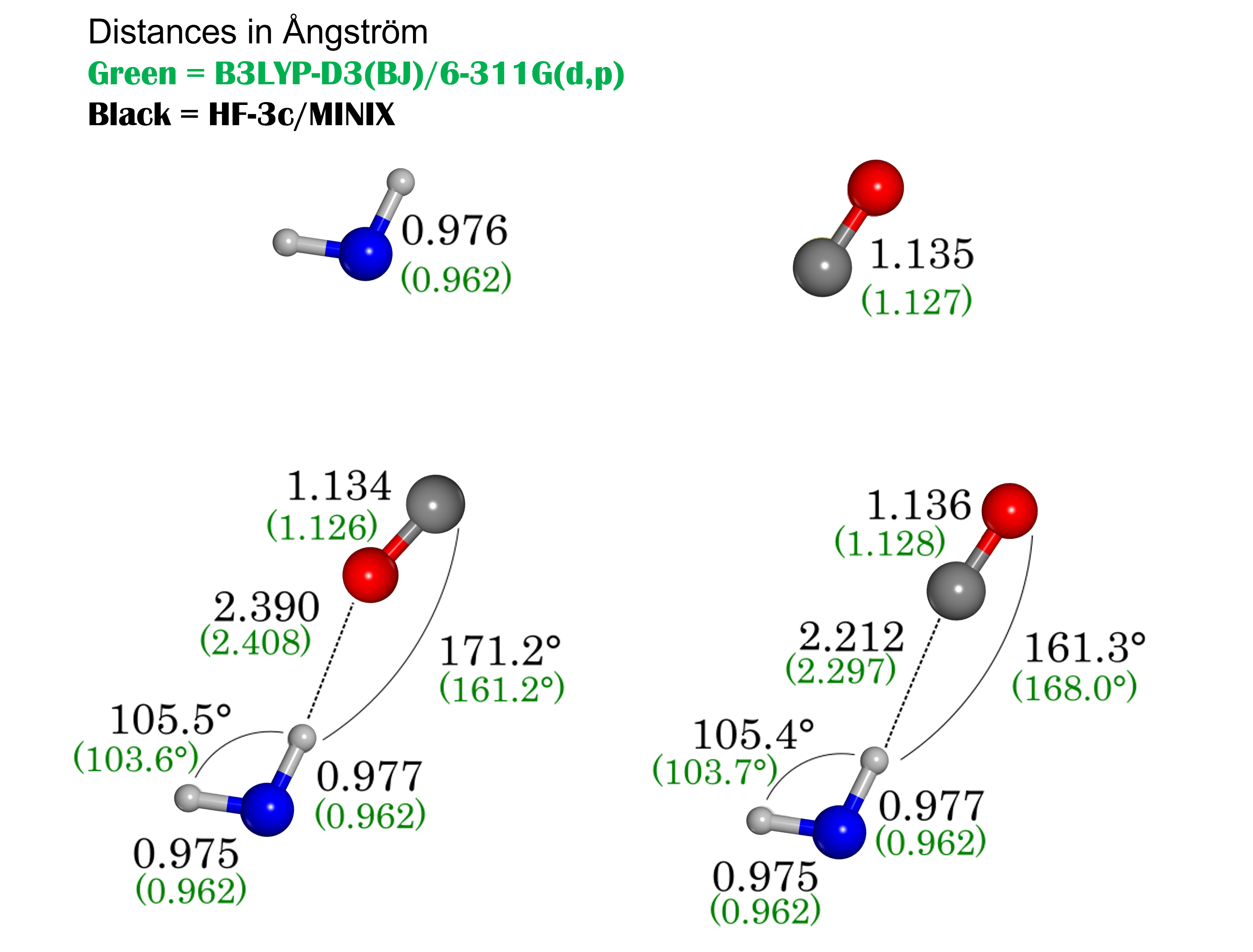}
    \caption{the interaction between CO and H$_2$O computed at HF-3c (black) and B3LYP-D3(BJ)/6-311G(d,p) (green) levels of theory. The H-bond established through the C-atom is energetically favoured over the one established through the O-atom of CO by only 0.3 kJ mol$^{-1}$ at HF-3c level against 3.9 kJ mol$^{-1}$ of B3LYP-D3(BJ).}
    \label{fig:dimer_h2o_co}
\end{figure}

\clearpage

\section{Reaction on (100) surface}
\begin{figure}
    \centering
    \includegraphics[width=0.75\columnwidth]{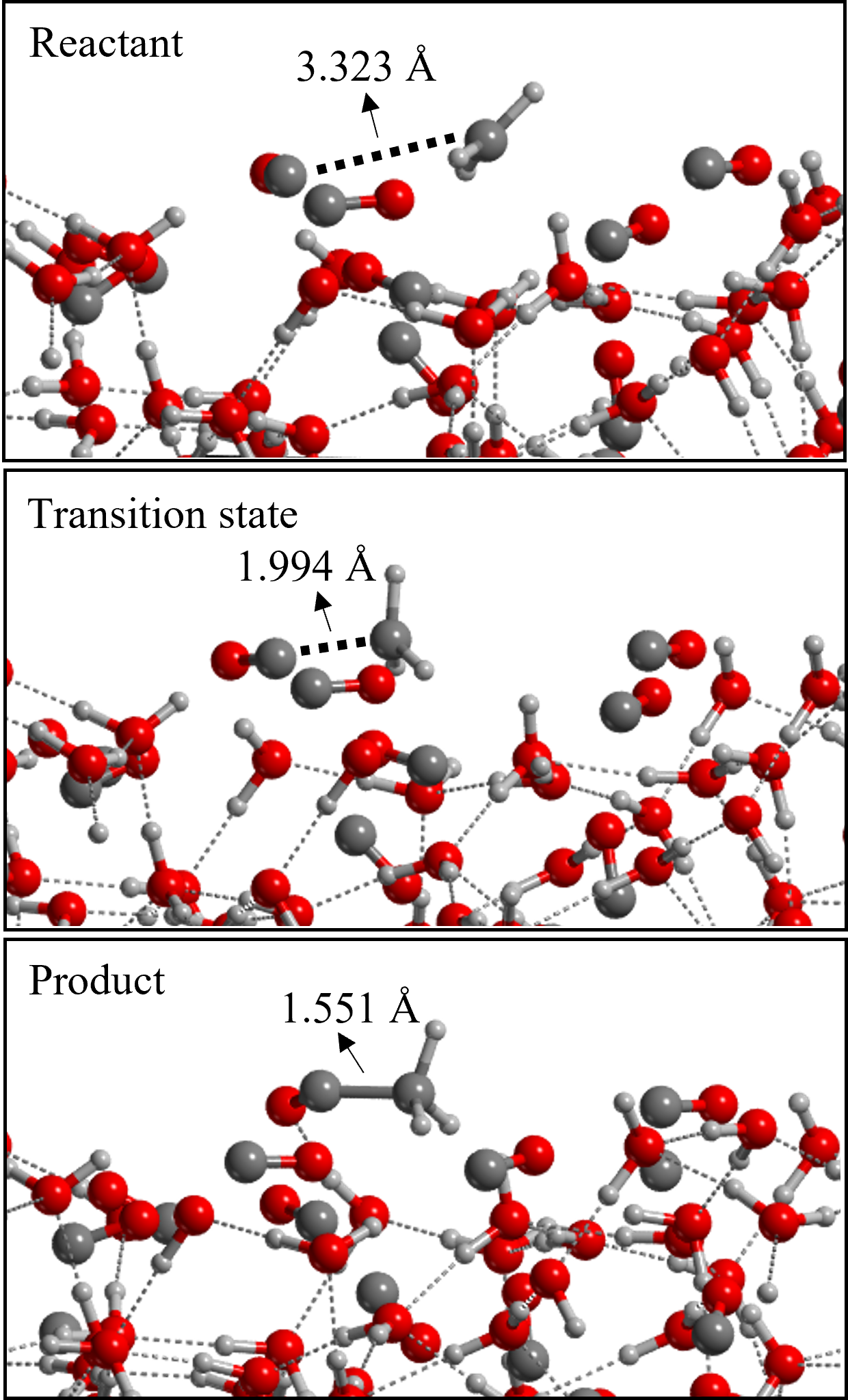}
    \caption{Reactant, transition state and product structures of the acetyl formation reaction on (100) surface, binding site \textbf{H1}. Color code: H, white; C, grey; O, red. }
    \label{fig:acetyl_formation}
\end{figure}

\begin{figure}
    \centering
    \includegraphics[width=0.75\columnwidth]{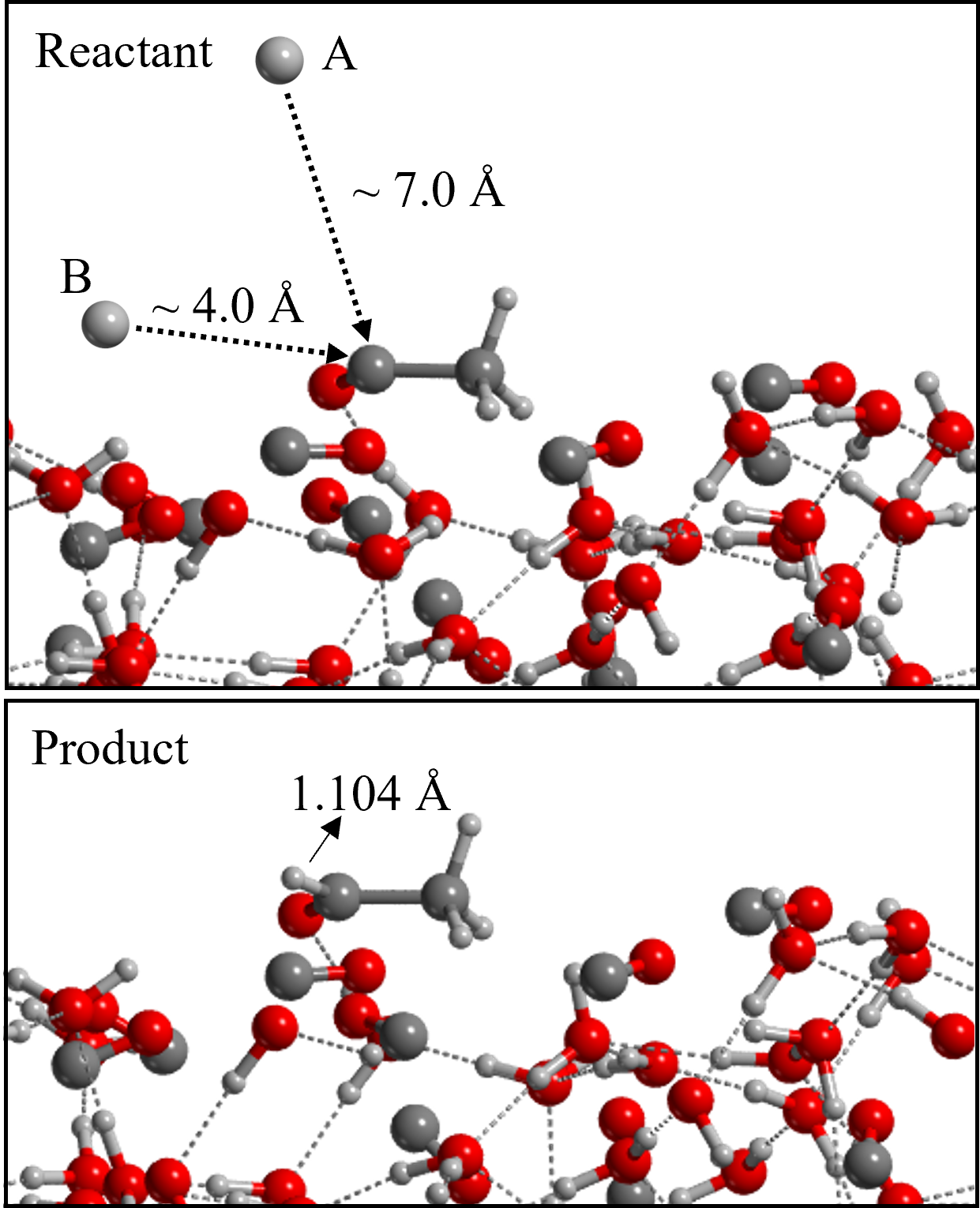}
    \caption{Reactants and product structures of the acetaldehyde formation reaction on (100) surface. The hydrogen atom approaching from the gas-phase (A) represents the Eley-Rideal mechanism, while in case of the Langmuir-Hinshelwood mechanism, the atom is adsorbed on the surface (B). Color code: H, white; C, grey; O, red.}
    \label{fig:acetaldehyde_formation}
\end{figure}

%%%%%%%%%%%%%%%%%%%%%%%%%%%%%%%%%%%%%%%%%%%%%%%%%%

% Don't change these lines
\bsp	% typesetting comment
\label{lastpage}
\end{document}